\documentclass[a4paper,11pt]{article}
\pdfoutput=1

\usepackage{jheppub}

\usepackage{hyperref}
\usepackage{graphicx}
\usepackage{amsmath}
\usepackage{xcolor}
\usepackage{amssymb}
\usepackage[utf8]{inputenc}
\usepackage[T1]{fontenc}
\usepackage{lmodern}
\newcommand{\eq}[1]{eq.~\eqref{eq:#1}}
\newcommand{\eqs}[2]{eqs.~\eqref{eq:#1} and \eqref{eq:#2}}
\newcommand{\eqsto}[2]{eqs.~\eqref{eq:#1}\,-\,\eqref{eq:#2}}
\newcommand{\eqss}[3]{eqs.~\eqref{eq:#1}, \eqref{eq:#2}, and \eqref{eq:#3}}
\renewcommand{\sec}[1]{sec.~\ref{sec:#1}}

\newcommand{\subsec}[1]{sec.~\ref{subsec:#1}}

\newcommand{\fig}[1]{fig.~\ref{fig:#1}}

\newcommand{\app}[1]{app.~\ref{app:#1}}

\newcommand{\rcites}[1]{refs.~\cite{#1}}
\newcommand{\rcite}[1]{ref.~\cite{#1}}

\newcommand{\ord}[1]{{\mathcal O}(#1)}

\newcommand{\ORD}[1]{{\mathcal O}\biggl(#1\biggr)}

\newcommand{\bare}{\mathrm{bare}}

\newcommand{\df}{\mathrm{d}}

\newcommand{\eps}{\epsilon}

\newcommand{\nn}{\nonumber}

\newcommand{\cD}{{\mathcal D}}
\newcommand{\cL}{{\mathcal L}}

\newcommand{\cusp}{\mathrm{cusp}}

\newcommand{\msb}{{\overline{\rm MS}}}
\newcommand{\as}{\alpha_s}

\newcommand{\zero}{{(0)}}

\graphicspath{ {figs/} }

\DeclareMathOperator{\tr}{tr}
\DeclareMathOperator{\bravac}{\big\langle  0 \big|}
\DeclareMathOperator{\ketvac}{\big|  0 \big\rangle}
\newcommand{\brab}{\big\langle  b_v \big|}
\newcommand{\ketb}{\big|  b_v \big\rangle}
\newcommand{\braZ}{\big\langle  Z \big|}
\newcommand{\ketZ}{\big|  Z \big\rangle}

\newcommand{\ketbin}{\big|  b_v,\mathrm{in} \big\rangle}
\newcommand{\brabout}{\big\langle  b_v, \mathrm{out} \big|}
\newcommand{\braboutzero}{\big\langle  b_v^\zero, \mathrm{out} \big|}

\newcommand{\ketbinzero}{\big|  b_v^\zero,\mathrm{in} \big\rangle}
\newcommand{\brabzero}{\big\langle  b_v^\zero \big|}
\newcommand{\ketbzero}{\big|  b_v^\zero \big\rangle}
\DeclareMathOperator{\Tbar}{\overline T}
\DeclareMathOperator{\T}{T}
\newcommand{\sumintZ}{\sum \!\!\!\!\!\!\!\! \int\limits_{Z}}
\newcommand{\hvbar}{\overline{h}_v}

\allowdisplaybreaks[2]

\preprint{\begin{flushright}
MITP/19-073 \\
P3H-19-042 \\
SI-HEP-2019-15
\end{flushright}}

\title{
Three-loop soft function for heavy-to-light quark decays}

\author[a,b]{Robin Br\"user,}
\emailAdd{brueser@uni-mainz.de}
\author[a,c]{Ze Long Liu,}
\emailAdd{liu@uni-mainz.de}
\author[a]{Maximilian Stahlhofen}
\emailAdd{mastahlh@uni-mainz.de}

\affiliation[a]{PRISMA$^+$ Cluster of Excellence, %Institute of Physics, 
Johannes Gutenberg University, D-55128 Mainz, Germany
}
\affiliation[b]{Theoretische Physik 1, Naturwissenschaftlich-Technische 
Fakult{\"a}t, Universit{\"a}t Siegen,
Walter-Flex-Straße 3, D-57068 Siegen, Germany
}
\affiliation[c]{Theoretical Division, Los Alamos National Laboratory, Los 
Alamos, New Mexico 87545, USA}

\abstract{
We compute the 1-jettiness soft function for the decay of a heavy quark into a 
light quark jet plus colorless particles at three-loop order in 
soft-collinear effective theory.
The 1-jettiness measurement fixes the total small light-cone momentum component 
of the soft radiation with respect to the jet direction. This soft function is 
a universal ingredient to the factorization of heavy-to-light quark decays in 
the limit of small 1-jettiness.
Our three-loop result is required for resummation at the 
N$^3$LL$^\prime$ level, e.g.\ near the endpoint in the photon energy
spectrum of the $B \to X_s \gamma$ decay.
It is also a necessary ingredient for future calculations of fully-differential 
heavy-to-light quark decay rates at N$^3$LO using the $N$-jettiness subtraction 
method, e.g.\ for semileptonic top decays.
Using our result for the soft anomalous dimension we confirm predictions on the 
universal infrared structure of QCD scattering amplitudes with a massive 
external quark at three loops.
}
\begin{document}
\maketitle

%%%%%%%%%%%%%%%%%%%%%%%%%%%%%%%%%%%%%%%%%%%%%%%%%%%%%%%%%%%%%%%%%%%%%%%%%%%%%%%%
\section{Introduction}
\label{sec:Intro}
%%%%%%%%%%%%%%%%%%%%%%%%%%%%%%%%%%%%%%%%%%%%%%%%%%%%%%%%%%%%%%%%%%%%%%%%%%%%%%%%

Heavy quark (bottom, top) decays are phenomenologically important Standard 
Model (SM) processes.
A prominent example is the decay $b \to s \gamma$.
In the SM it is loop-suppressed and represents a promising window to 
new physics. In particular, beyond SM interactions due to flavor changing 
neutral currents could add a measurable effect on the decay rate of 
$B \to X_s \gamma$ on top of the small SM background.
In the phenomenologically relevant region of large photon energies the decay 
rate $\Gamma(B\to X_s \gamma)$ factorizes as~\cite{Korchemsky:1994jb}
\begin{align}
\frac{\df \Gamma}{\df E_\gamma} =
H(E_\gamma,m_b,\mu)  \int \! \df \omega \,m_b\, J(m_b\, \omega,\mu)\,
\mathcal{S}(\Delta - \omega, \mu)
+\ORD{\frac{\Delta}{m_b},\frac{\Lambda_\mathrm{QCD}}{m_b} }
\,,
\label{eq:fact}
\end{align}
where $\Delta=m_b-2E_\gamma$.%
\footnote{For brevity we have absorbed a constant overall factor including 
electroweak and electromagnetic couplings as well as CKM matrix elements in the 
hard function $H$.}
Within soft-collinear effective theory (SCET)~\cite{Bauer:2000ew, Bauer:2000yr, 
Bauer:2001ct, Bauer:2001yt, Bauer:2002nz, Beneke:2002ph} this factorization 
theorem was proven in \rcite{Bauer:2001yt}.
The hard function $H$ encodes the short distance (electroweak) interaction 
and its virtual quantum corrections at and beyond the hard scale $m_b \sim 
E_\gamma$. Explicit expressions up to two loops can be found in 
\rcites{Ligeti:2008ac,Ali:2007sj}.
The jet function $J$ describes the collinear radiation in the final 
state jet initiated by the (massless) $s$ quark and is governed by the 
virtuality scale $\sqrt{m_b\Delta}$. In \rcite{Bruser:2018rad} we 
computed the massless quark jet function to three-loop order (see also 
\rcite{Banerjee:2018ozf}).
Finally, $\mathcal{S}$ denotes the $B$-meson shape 
function~\cite{Neubert:1993um,Bigi:1993ex} which describes the physics at scales
smaller or similar to $\Delta$. For $\Delta \gtrsim \Lambda_\mathrm{QCD}$
nonperturbative effects are sizable and one can further factorize 
$\mathcal{S}$ into a purely perturbative `heavy-to-light soft function' 
$S_\mathrm{hl}$ and a nonperturbative shape function $F$~\cite{Ligeti:2008ac}:
\begin{align}
\mathcal{S}(\omega,\mu) = \int \! \df \omega^\prime \, 
S_\mathrm{hl} (\omega -\omega^\prime, \mu) \,F(\omega^\prime) %+m_b-m_B)
\,,
\label{eq:Sfact}
\end{align}
with $\int \! \df \omega\, F(\omega) = 1$.
The soft function $S_\mathrm{hl}$ can be expressed as a partonic $b$-quark 
matrix 
element, see \eq{pertshapefctdef}, and was computed to two-loop order in 
\rcite{Becher:2005pd}.
The functions $J$, $S_\mathrm{hl}$, and $F$ vanish when their first argument is 
negative. This entails finite integration ranges in \eqs{fact}{Sfact}. 
The perturbative factorization functions $H$, $J$, and $S_\mathrm{hl}$ depend 
individually on the common renormalization scale $\mu$, while the
total decay rate in \eq{fact} is $\mu$ independent to the 
perturbative order one is working at.
The renormalization group (RG) evolution of the hard, jet, and, soft functions 
is therefore not independent, but subject to a consistency relation (which 
will be relevant later).
The combined RG running of the different functions to the common scale $\mu$ 
eventually resums large logarithms of the ratios between the hard, jet, and 
soft (matching) scales $\mu_H \sim m_b$, $\mu_J \sim \sqrt{m_b\Delta}$, and 
$\mu_S \sim \Delta$.

A factorization theorem analogous to \eq{fact} also holds for the decay 
$B\to X_u \ell \bar\nu$.
In particular the involved jet and shape functions are the same.
The nonperturbative function $F$ can thus be obtained from a fit to 
experimental data for the differential spectrum of one or both decays, see 
e.g.\ \rcite{Bernlochner:2013gla}, and then be used for 
theoretical predictions. For details we refer to \rcite{Ligeti:2008ac}.
The current state of the art for such predictions includes resummation 
at the primed next-to-next-to-leading logarithmic (NNLL$^\prime$) 
level~\cite{Ligeti:2008ac,Becher:2006pu}, where the NNLL 
expression~\cite{Neubert:2004dd} is augmented with the next-to-next-to-leading 
order (NNLO) corrections to the factorization functions at their matching 
scales, i.e.\ to $H(\mu_H$), $J(\mu_J)$, $S_\mathrm{hl}(\mu_S)$.%
\footnote{For details and advantages of the primed counting, see e.g.\
\rcite{Almeida:2014uva}.}
Still, the uncertainties from missing higher-order perturbative 
corrections represent a major contribution to the total error 
budget~\cite{Bernlochner:2013gla}.

In order to reach N$^3$LL$^\prime$ accuracy of the decay rate  in \eq{fact} the 
three-loop corrections to the 
hard, jet, and soft, functions along with their anomalous dimensions are 
required. Our three-loop calculation of the jet function in 
\rcite{Bruser:2018rad} represents a first step toward this goal. In the present 
paper we calculate the  soft function $S_\mathrm{hl}$ at 
three loops, while the three-loop hard function is left for future work.
We also give explicit expressions for all three-loop (noncusp) anomalous 
dimensions necessary for N$^3$LL$^{(\prime)}$ resummation in \eq{fact}.

Another possible application of our result is within the context of 
$N$-jettiness subtractions~\cite{Boughezal:2015dva,Gaunt:2015pea}.
In its simplest version the latter is an infrared (IR) slicing method which uses 
the observable $N$-jettiness $\mathcal{T}_N$~\cite{Stewart:2010tn} as an 
auxiliary resolution variable for soft and collinear real emissions.
It was employed amongst others to compute the fully-differential decay rate of 
the semileptonic top decay $t\to W^+(l^+\nu)b$ at NNLO in QCD~\cite{Gao:2012ja}.
In this case the resolution variable is $\mathcal{T}_1$ (1-jettiness).
For $\mathcal{T}_1 < \mathcal{T}_\mathrm{cut}$ the decay rate is given by a 
factorization formula analogous to \eq{fact}, but with 
$\mathcal{S}=S_\mathrm{hl}$, provided that 
$\mathcal{T}_\mathrm{cut}$ is small enough to neglect 
$\ord{\mathcal{T}_\mathrm{cut}/m_t}$ power corrections at the desired precision.
For $\mathcal{T}_1 > \mathcal{T}_\mathrm{cut}$ there is at least one additional 
hard parton in the final state. On the other hand quantum corrections to this 
part of the decay rate are only needed at one order lower in the perturbative 
expansion. In case of the NNLO $t\to W^+(l^+\nu)b$ decay the $\mathcal{T}_1 > 
\mathcal{T}_\mathrm{cut}$ piece can therefore be computed with standard 
(numerical) NLO technology.
The soft function $S_\mathrm{hl}$ we calculate in the present paper equals the 
soft function in the 1-jettiness factorization theorem for any heavy-to-light 
quark decay. The new three-loop contribution is thus a necessary 
ingredient for future N$^3$LO calculations of differential decay rates based on 
the $N$-jettiness method, not only for semileptonic top, but also any other 
heavy-to-light quark decay.

The outline of this paper is as follows.
In \sec{def} we give four (slightly) different definitions of the 
heavy-to-light soft function $S_\mathrm{hl}$ and show their equivalence.
We give details on our three-loop calculation based on one of these definitions 
in \sec{calc}. In \sec{res} we present our results for the renormalized soft 
function and its anomalous dimension. We also use the latter to check the 
universal infrared structure of QCD scattering amplitudes that have a massive 
quark leg.
We briefly summarize our findings in \sec{sum}.

%%%%%%%%%%%%%%%%%%%%%%%%%%%%%%%%%%%%%%%%%%%%%%%%%%%%%%%%%%%%%%%%%%%%%%%%%%%%%%%%
\section{Definitions}
\label{sec:def}
%%%%%%%%%%%%%%%%%%%%%%%%%%%%%%%%%%%%%%%%%%%%%%%%%%%%%%%%%%%%%%%%%%%%%%%%%%%%%%%%

The 1-jettiness soft function for heavy-to-light decays is defined by the 
vacuum matrix element
\begin{align}
 S_\mathrm{hl}(\omega) := \frac{1}{N_c} \tr \bravac  
\Tbar \Big[ \big(X_+\big)^{\!\dagger}\!(0) \, Y_-(0) \Big]
\,\delta(\omega - n\!\cdot\!\hat p) \,
\T  \Big[ \big(Y_-\big)^{\!\dagger}\!(0)\, X_+(0) \Big]
 \ketvac\,,
 \label{eq:Shldef}
\end{align}
with the soft momentum operator $\hat p^\mu$ and the Wilson lines
\begin{align}
 X_+ (x) &= \mathrm{P} \exp \Big[i g \! \int_{-\infty}^0 \!\!\! \df s \; 
  v\!\cdot\! A(x + sv) \Big]\,, \\
 Y_- (x) &= \overline {\mathrm{P}} \exp \Big[- i g \!\int^\infty_0 \!\!
  \df s \; n\!\cdot\! A(x + sn) \Big]\,,
\end{align}
where $A_\mu(x) \equiv A_\mu^a(x) T^a$ is the (ultra)soft 
SCET$_{(\mathrm{I})}$ gluon field, $v_\mu$ 
is the heavy quark velocity ($v^2=1$), $n_\mu$ is the light-like jet direction 
($n^2=0$) and $\mathrm{P}$ ($\overline {\mathrm{P}}$) denotes (anti-)path 
ordering of the $A_\mu$ including their $SU(N_c)$ color generators $T^a$.
The trace in \eq{Shldef} is over color indices, and $\T[\ldots]$ and 
$\Tbar[\ldots]$ represent time- and anti-time-ordered products of the field 
operators $A_\mu^a(x)$, respectively. 
The argument $\omega$ of $S_\mathrm{hl}$ can be regarded as the (appropriately 
normalized) soft contribution to the 1-jettiness observable 
$\mathcal{T}_1$, cf.\ \rcite{Stewart:2010tn}.

The soft function in \eq{Shldef} equals the perturbative contribution to the 
shape function in \eq{Sfact}:%
\footnote{In the following we take the decaying heavy quark without loss of 
generality to be a bottom quark.}
\begin{align}
S_\mathrm{hl}(\omega) =  \brab \hvbar(0) \, \delta(\omega + i n\! \cdot \!D)\, 
h_v(0) \ketb\,,
 \label{eq:pertshapefctdef}
\end{align}
where averaging over color and spin of the external HQET $b$-quark states is 
understood, the latter are normalized such that $\langle b_v| \hvbar(0) h_v(0) 
|b_v\rangle=1$, $h_v$ is the HQET heavy quark field with velocity $v$, and 
$D^\mu = \partial^\mu + i g A^\mu(0)$.
This $b$-quark matrix element was calculated to $\ord{\alpha_s}$ in 
\rcites{Bauer:2003pi,Bosch:2004th} and $\ord{\alpha_s^2}$ in 
\rcite{Becher:2005pd}.
The equivalence to \eq{Shldef} can be seen as follows:
\begin{align}
 &\brab \hvbar(0) \, \delta(\omega + i n\! \cdot \!D)\, h_v(0) \ketb \nn \\
 &= \brab \hvbar(0) \,Y_-(0)\, \delta(\omega + i n\! \cdot \!\partial)\,
 \big(Y_-\big)^{\!\dagger}\!(0)\, h_v(0) \ketb\\
 &= \brab \Tbar \Big[ \hvbar(0) \,Y_-(0) \Big]\, 
 \delta(\omega - n\!\cdot\!\hat p) \,\T \Big[ \big(Y_-\big)^{\!\dagger}\!(0)\, 
h_v(0) \Big] \ketb \label{eq:equiv2}\\
 &= \brabzero \hvbar^\zero(0) \Tbar \Big[ \big(X_+\big)^{\!\dagger}\!(0)
 \,Y_-(0) \Big]\, \delta(\omega - n\!\cdot\!\hat p) \,\T \Big[ 
\big(Y_-\big)^{\!\dagger}\!(0)\, X_+(0) \Big] h_v^\zero(0) \ketbzero 
\label{eq:equiv3}\\
 &= S_\mathrm{hl}(\omega)\,.
\end{align}
In \eq{equiv2} we could introduce the $\T$ and $\Tbar$ symbols, because the 
field operators in $Y_-$ are already anti-time-ordered by default as a 
consequence of the anti-path-ordering, and Hermitian conjugation reverses the 
order. Similarly the $\T$ and $\Tbar$ symbols in \eqs{equiv3}{Shldef} are in 
fact redundant, but kept for clarity.
In \eq{equiv3} we performed the HQET field redefinition
\begin{align}
 h_v(x) \to X_+(x) \, h_v^\zero(x)\,,
 \label{eq:fieldredef}
\end{align}
where the new (sterile) field $h_v^\zero$ does not interact with soft gluons 
anymore, see e.g.\ \rcite{Bauer:2001yt}.
Note that given the (anti)-time-ordering the external $b$-quark states should 
be 
interpreted as  `in' states,
$|b_v\rangle = |b_v , \mathrm{in}\rangle$, 
$\langle b_v| = \langle b_v,\mathrm{in}|$. 
The LSZ reduction formula relates the asymptotic `in'/`out' $b$-quark states of 
the S-matrix element to weighted integrals over interpolating 
$\overline{h}_v(x)$/$h_v(x)$ field operators acting on the vacuum at macroscopically 
large negative/positive times. 
Performing the field redefinition in \eq{fieldredef} one also has to take into 
account factors of $X_+$ from these interpolating fields. 
For the `in' states this factor is trivial and we can effectively 
replace~\cite{Arnesen:2005nk}
\begin{align}
\ketbin_i \to \Big[\big(X_+\big)^{\!\dagger}\!(t\!=\!-\infty, \vec x=0)\Big]_{ji}\,
\ketbinzero_j = \ketbinzero_i\;,
\label{eq:fieldredefin}
\end{align}
where $i$, $j$ are color indices in the fundamental representation. 
Here and in the following we assume the $b$ quark to be at rest, 
i.e.\ $v^\mu=(1,\vec 0)$ for simplicity.
The spatial position $\vec x$ of the endpoint of the Wilson line $X_+$ in 
\eq{fieldredefin} is fixed by the position of the field operator, 
here $h_v(0)$, acting on the asymptotic state, because 
$\langle 0|h_v(x) \overline{h}_v(y)|0\rangle \propto \delta^{(3)}(\vec x - \vec y)$.
Finally, the sterile HQET quark field operators in \eq{equiv3} annihilate the sterile 
external quarks and the color averaging implicit in the $b$-quark matrix 
elements 
translates to $1/N_c$ times the color trace in \eq{Shldef}.

For the actual calculation of the soft function it is convenient to express it 
as the imaginary part (discontinuity) of a ($1 \to 1$) `forward scattering' 
matrix element.
Starting from \eq{equiv2} and inserting a complete set of states we have
\begin{align}
 &S_\mathrm{hl}(\omega) = 
 \sumintZ  \; \delta(\omega - p_Z^+) \, \Big| \braZ \T 
 \Big[ \big(Y_-\big)^{\!\dagger}\!(0)\, h_v(0) \Big] \ketb  \Big|^2 \nn \\
 &\quad= \mathrm{Im}\bigg[\, i \sumintZ\, 
 \int_{-\infty}^\infty \!\frac{\df s}{2\pi} \,\frac{i}{s + i0} \;
 \delta\Big(\frac{\omega}{2} - \frac{p_Z^+}{2}-s \Big) \,
 \Big| \braZ \big(Y_-\big)^{\!\dagger}\!(0)\, h_v(0) \ketb  \Big|^2 \;\bigg] \\
 &\quad= \mathrm{Im}\bigg[\, i \sumintZ\,  
 \int_{-\infty}^\infty \!\frac{\df x^-}{2\pi}\,e^{\frac{i}{2} \omega x^-} \!\!
 \int_{-\infty}^\infty \! \frac{\df s}{2\pi} \,\frac{i e^{-i s x^-}}{s + i0} 
 \nn\\
 &\qquad \qquad
 \times \brab e^{\frac{i}{2} \hat p^+ x^-}\, \overline{ h}_v(0)\,Y_-(0) 
 \,e^{-\frac{i}{2} \hat p^+ x^-} \ketZ \braZ \big(Y_-\big)^{\!\dagger}\!(0)\, 
h_v(0) \ketb \bigg] \label{eq:equivdis2} \\
 &\quad= \mathrm{Im}\bigg[\,i  \int_{-\infty}^\infty \!\frac{\df x^-}{2\pi}
 \,e^{\frac{i}{2} \omega x^-} \theta(x^-)
 \brab \hvbar \big(x^- {\textstyle \frac{n}{2}}\big)\,Y_- 
 \big(x^- {\textstyle \frac{n}{2}} \big) \,\big(Y_-\big)^{\!\dagger}\!(0)\, 
h_v(0) \ketb \bigg]  
 \label{eq:equivdis3}\\
 &\quad= \mathrm{Im}\bigg[\,i  \int_{0}^\infty \!\frac{\df x^-}{2\pi}\,
 e^{\frac{i}{2} \omega x^-} 
  \brab \T \Big[ \hvbar \big(x^- {\textstyle \frac{n}{2}}\big)\, 
  \mathrm{P} \exp \Big[i g \!\int_{0}^{x^-\!/2} \!\! \df s \; n\!\cdot\! A(sn) 
  \Big] \, h_v(0) \Big] \ketb \bigg] \,.
  \label{eq:dispersiveb}
\end{align}
Here we use the usual light-cone (Sudakov) decomposition of four vectors: 
$a^\mu = a^-n^\mu/2 + a^+\bar n^\mu/2 + a^\mu_\perp$ with $n^2=\bar n^2=0$ and 
$\bar n \cdot n=2$.
Note that the momentum operator $\hat p^+$ acting to the left on the external 
HQET state in \eq{equivdis2} vanishes, because the external heavy quarks are 
onshell and therefore have zero residual (soft) four-momentum.
While for space-like distance $x^2<0$ the field operators commute, the theta 
function $\theta(x^-)$ in \eq{equivdis3} implies $t>0$ for time-like distances 
$x^2>0$. After combining the two Wilson lines using their unitarity property 
the 
remaining field operators are therefore automatically time-ordered. To make 
this 
manifest we explicitly inserted the $\T$ symbol in \eq{dispersiveb}.

\begin{figure*}[t]
	\begin{center}
		\includegraphics[width=0.4\textwidth]{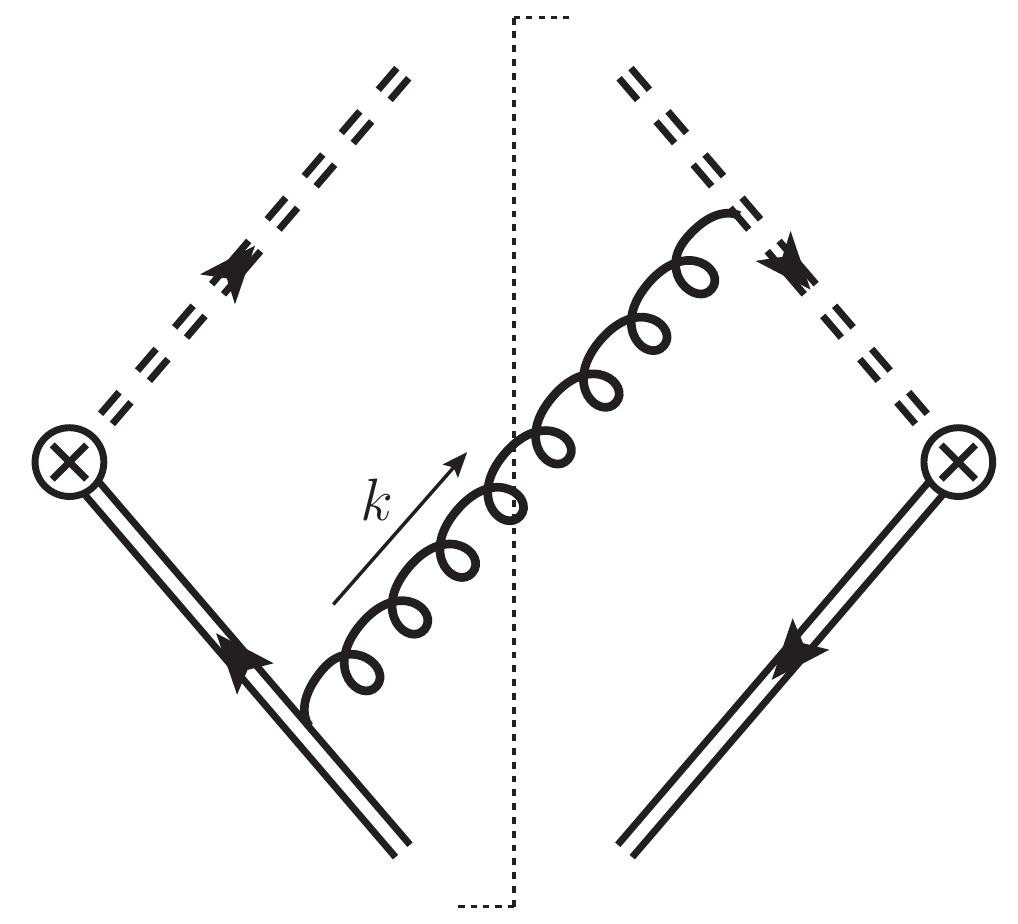}
		\put(-110,120){$\delta(\omega\!-\!k^+)$}
		\raisebox{16 ex}{\;\;=\quad $\displaystyle 
		\mathrm{Im} \Bigg[\, \frac{i}{\pi} \times$}
		\hspace{-2.5 ex}
		\raisebox{9.6 ex}{
			\includegraphics[width=0.4\textwidth]{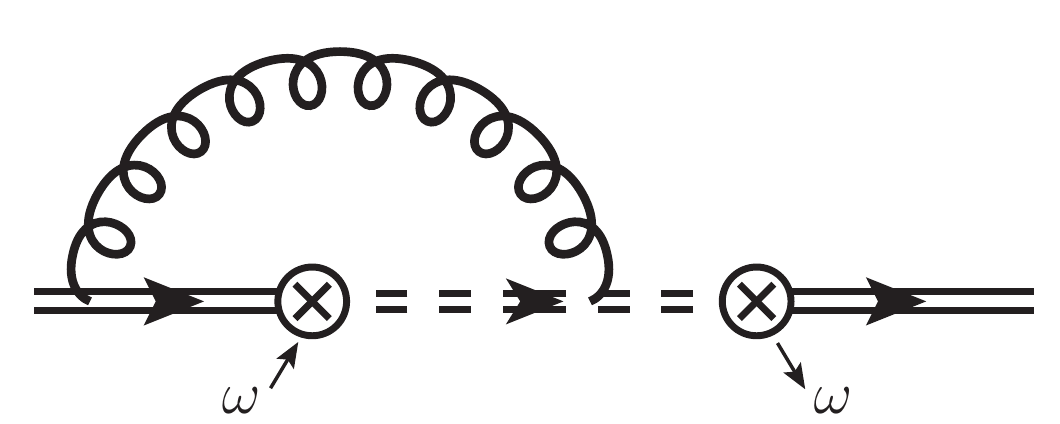}
		}
		\hspace{-2 ex}
		\raisebox{16 ex}{$\Bigg]$}
	\end{center}
	\vspace{-2 ex}
\caption{The cut diagram on the left hand side represents one 
$\ord{\alpha_s$} contribution to $S_\mathrm{hl}$ in \eq{Shldef}. The phase 
space 
integration of the soft gluon (with outgoing momentum $k^\mu$) crossing the 
final state cut is restricted by the measurement function $\delta(\omega - 
n\!\cdot\!\hat p)$.
The equivalent contribution to the expression in \eq{dispersiveWL} comes from 
the loop diagram on the right hand side, where the external lightcone momentum 
$\omega$ is routed through the light-like Wilson line as indicated. Here and in 
the following dashed double lines represent light-like Wilson lines ($Y$) and 
solid double lines represent time-like Wilson lines ($X$). The arrows on the 
Wilson lines indicate the fermion flow of the respective original quarks, which 
coincides with the direction of the path as well as the time ordering in this 
case.
\label{fig:cutIm}}
\end{figure*}

We can now again perform the field redefinition in \eq{fieldredef}.
This time, however, the time ordering in \eq{dispersiveb} implies that 
$|b_v\rangle = |b_v , \mathrm{in}\rangle$ and 
$\langle b_v| = \langle b_v,\mathrm{out}|$.
In contrast to \eq{fieldredefin} the field redefinition now induces a non-trivial 
factor from the interpolating fields generating the `out' 
state~\cite{Arnesen:2005nk}:
\begin{align}
{\phantom{\rangle}}_i\brabout \to {\phantom{\rangle}}_j\braboutzero
\Big[X_+(t\!=\!+\infty,x^- \textstyle \frac{\vec n}{2})\Big]_{ij} = 
{\phantom{\rangle}}_j\braboutzero 
\Big[ \big(X_-\big)^{\!\dagger}\!\big(x^- {\textstyle \frac{n}{2}}\big) 
\,X_+\big(x^- {\textstyle \frac{n}{2}}\big)\Big]_{ij}
\;.
\end{align}
Using $r=x^-/2$ as integration variable we thus obtain
\begin{align}
S_\mathrm{hl}(\omega) = 
\mathrm{Im}\bigg[\,\frac{i}{\pi}\, \frac{1}{N_c} \tr
\int_{0}^\infty \!\!\df r \,e^{i \omega r} 
\bravac \T \Big[ \big(X_-\big)^{\!\dagger}\!(rn)\, 
\mathrm{P} \exp \Big[i g \!\int_{0}^{r} \!\! \df s \; n\!\cdot\! A(sn) \Big] 
\, X_+(0) \Big] \ketvac \bigg] \,.
\label{eq:dispersiveWL}
\end{align}
In this expression the time-like Wilson lines extend from $t=-\infty$ to $t=0$
(incoming) and from $t=r$ to $t=+\infty$ (outgoing), respectively. 
The light-like Wilson line connects the points $0$ and $rn^\mu$. 
The Wilson line correlator in \eq{dispersiveWL} 
can be straightforwardly evaluated in terms of (momentum-space) Feynman diagrams 
using the usual Feynman rules for Wilson lines in QCD. 
The equivalence of \eqs{Shldef}{dispersiveWL} is illustrated on the 
diagrammatic level at one loop in \fig{cutIm}.
In \fig{diags} we show some examples of corresponding three-loop diagrams.
At $\ord{g^0}$ (tree level) $S_\mathrm{hl}$ is, according to \eq{dispersiveWL}, 
proportional to the discontinuity of a single light-like Wilson line propagator 
with soft light-cone momentum $\omega$:
\begin{align}
S_\mathrm{hl}^\zero(\omega) =
\mathrm{Im}\bigg[\,\frac{i}{\pi}\, \int_{-\infty}^\infty 
\!\!\df r \,e^{i \omega r} \theta(r) \bigg] = 
\mathrm{Im}\bigg[\,\frac{i}{\pi}\,\frac{i}{\omega + i0} \,\bigg] 
= \delta(\omega) \,.
\end{align}

To conclude this section we comment on the relation of $S_\mathrm{hl}$ to the 
analogous 1-jettiness soft functions where one or both Wilson lines in 
\eq{Shldef} are changed from incoming to outgoing or vice versa.
In the underlying full QCD processes the external heavy and light quark lines 
are correspondingly crossed from initial to final state or vice versa.
Some of these soft functions are e.g.\ relevant for $s$- and $t$-channel single 
top production as well as charm production in deep-inelastic neutrino scattering 
(`light-to-heavy DIS'). 
For state-of-the-art fully-differential NNLO predictions we 
refer to \rcite{Liu:2018gxa}, \rcites{Berger:2016oht,Berger:2017zof}, and 
\rcite{Berger:2016inr}, respectively.
The soft function for the light-to-heavy DIS process is for instance simply 
given by interchanging $X \leftrightarrow Y$ (i.e.\ $v^\mu \leftrightarrow 
n^\mu$ in the Wilson lines) in \eq{Shldef}.
Up to two loops the soft functions for the crossed processes can be shown to 
equal $S_\mathrm{hl}$ as defined in \eq{Shldef} in analogy to the massless 
case~\cite{Kang:2015moa}.
Unfortunately there is, to the best of our knowledge, no simple argument why 
this equality should hold at three loops and beyond, not even between
heavy-to-light decay and light-to-heavy DIS soft functions.%
\footnote{In \rcite{Moult:2018jzp} an all-order proof for the equality of two  
transverse momentum dependent soft functions, one with incoming, one with 
outgoing oppositely directed light-like Wilson lines based on time reversal 
symmetry of the vacuum is given. An analogous proof can however not be provided 
for our case because gluon field operators from the time-like and light-like 
Wilson lines do not commute.}
A dedicated three-loop analysis along the lines of \rcite{Kang:2015moa} would 
require to derive the analytic structure for the relevant two-loop 
single-emission and one-loop double-emission heavy-light soft currents, which 
is beyond the scope of this work.

%%%%%%%%%%%%%%%%%%%%%%%%%%%%%%%%%%%%%%%%%%%%%%%%%%%%%%%%%%%%%%%%%%%%%%%%%%%%%%%%
\section{Calculation}
\label{sec:calc}
%%%%%%%%%%%%%%%%%%%%%%%%%%%%%%%%%%%%%%%%%%%%%%%%%%%%%%%%%%%%%%%%%%%%%%%%%%%%%%%%

\begin{figure*}[t]
  \begin{center}
	\includegraphics[width=0.27\textwidth]{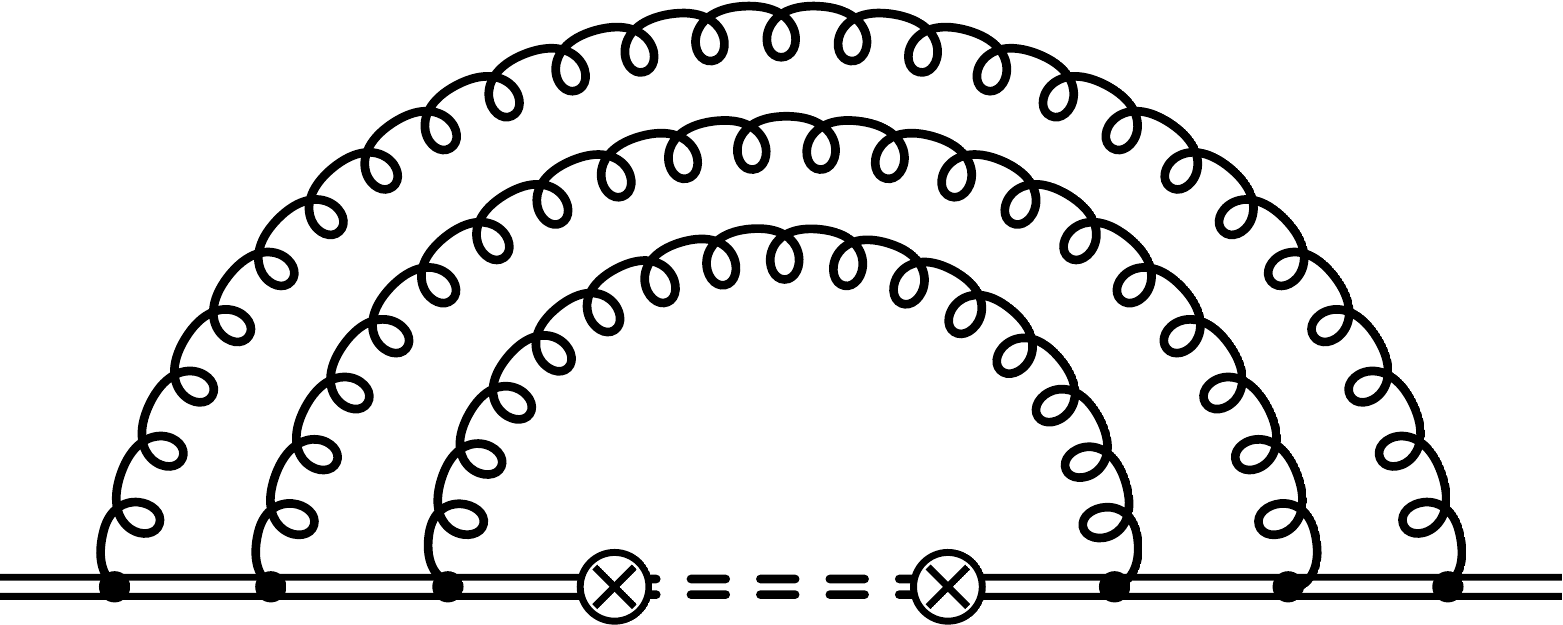}
	\hspace{0.09in}
	\includegraphics[width=0.27\textwidth]{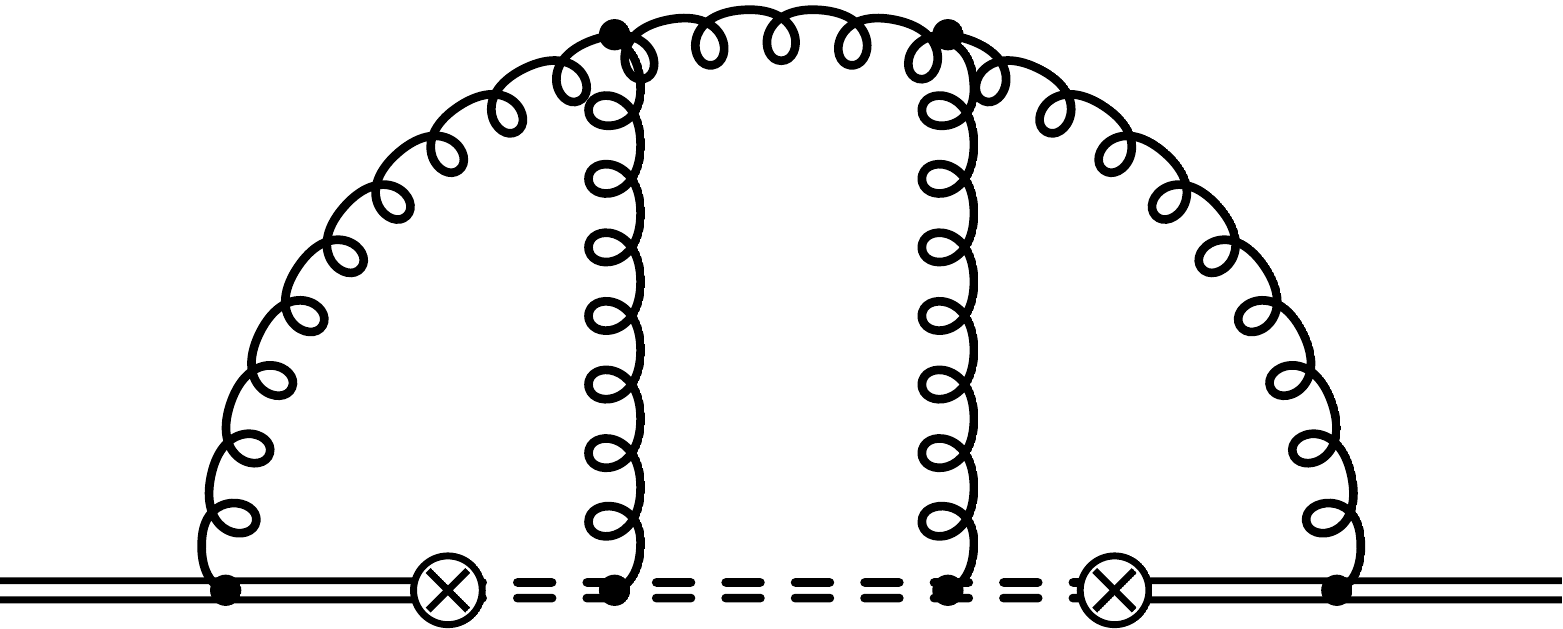}
	\hspace{0.09in}
	\includegraphics[width=0.27\textwidth]{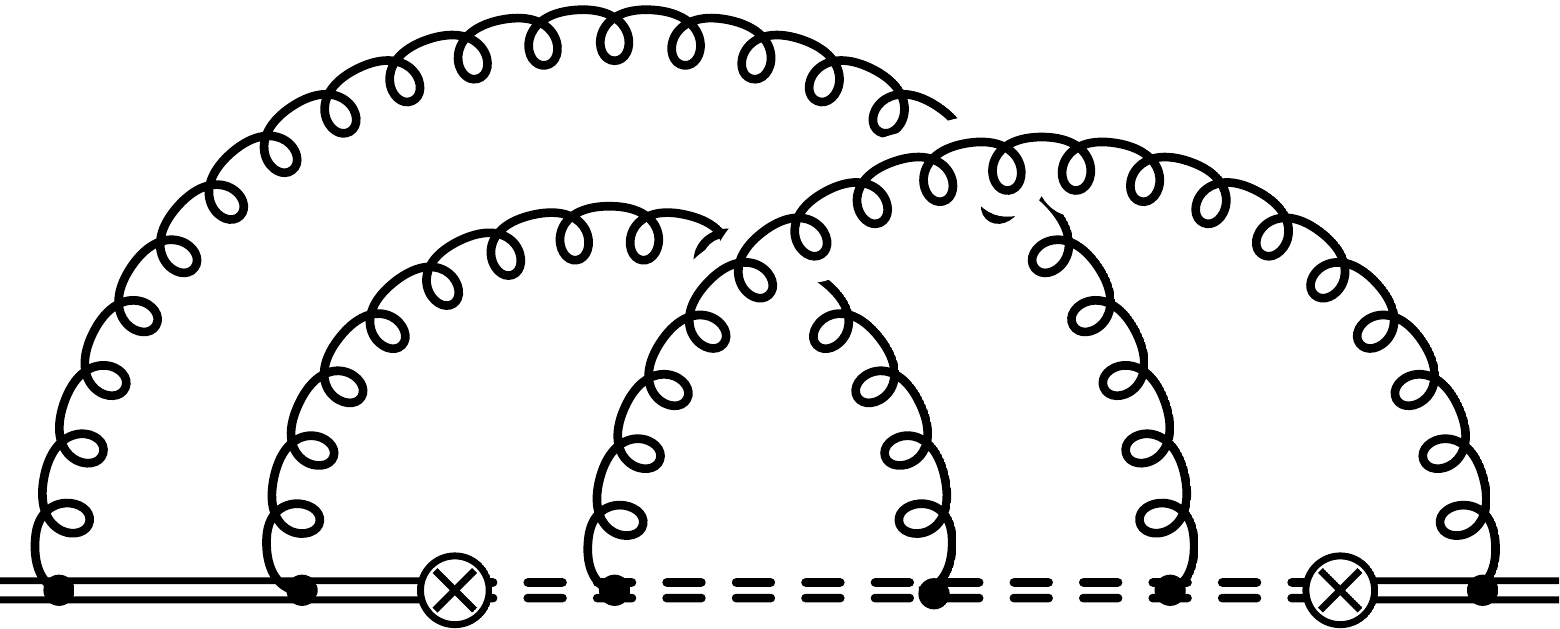}
	\\
	\vspace{0.19in}
	\includegraphics[width=0.27\textwidth]{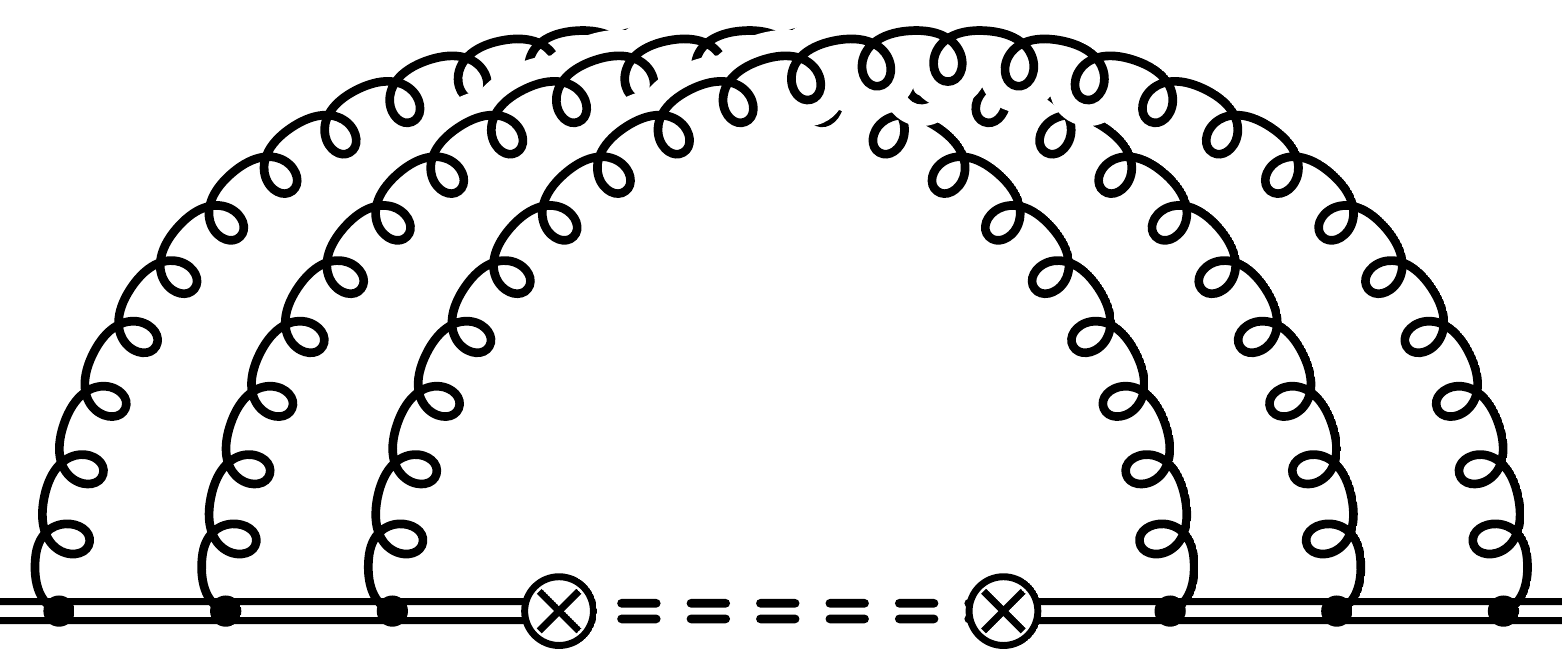}
	\hspace{0.09in}
	\includegraphics[width=0.27\textwidth]{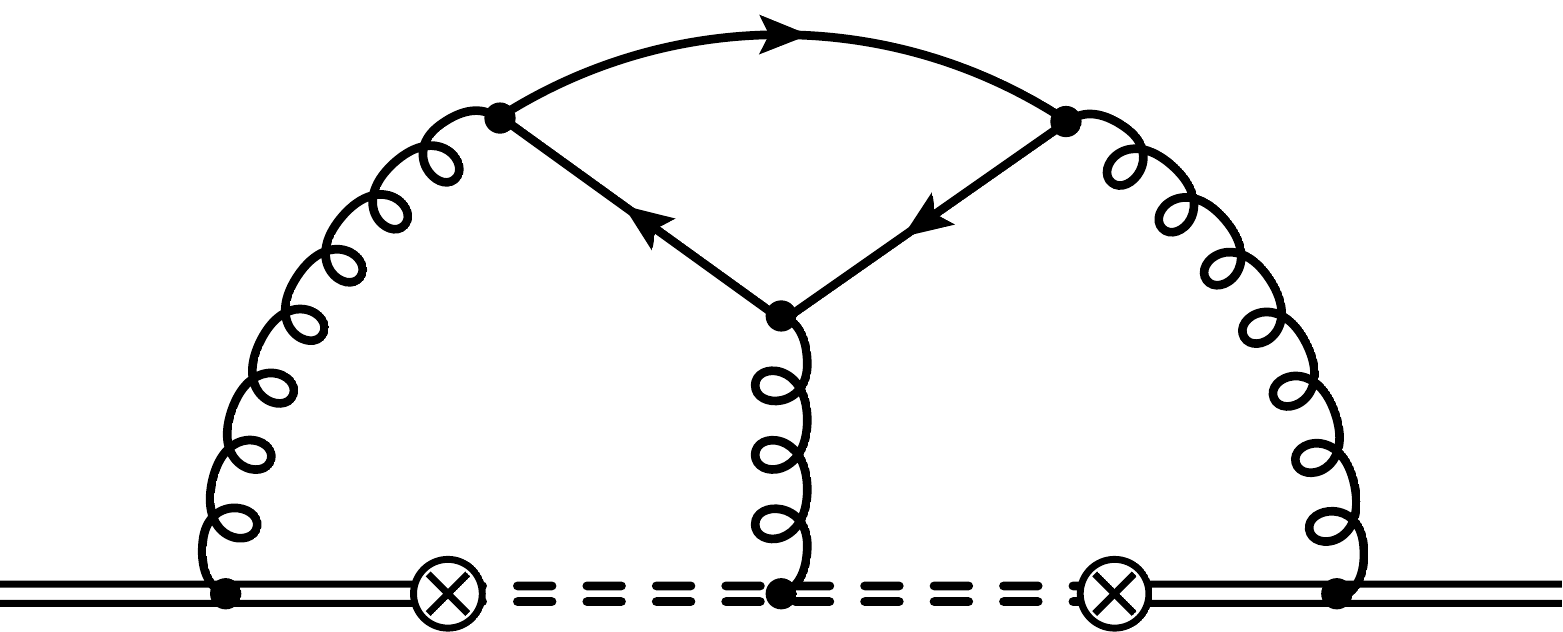}
	\hspace{0.09in}
	\includegraphics[width=0.27\textwidth]{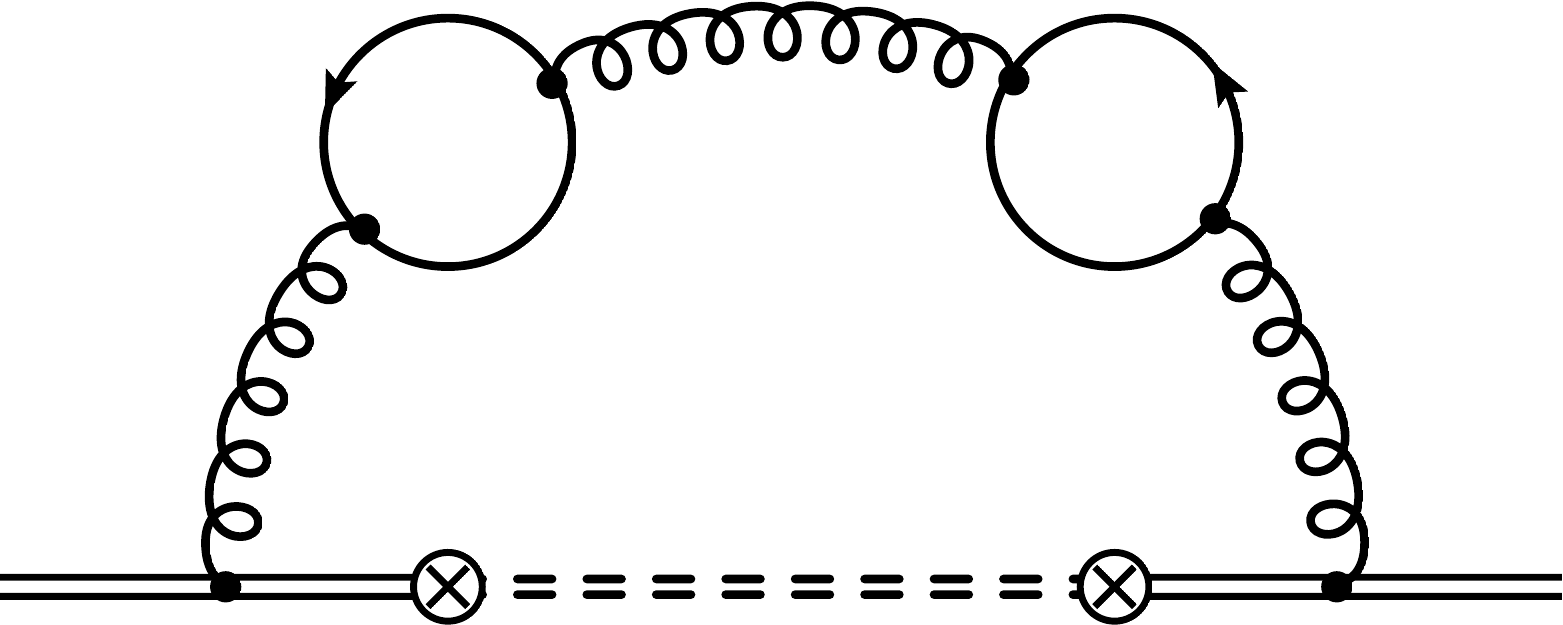}
  \end{center}
\vspace{2ex}
\caption{Sample Feynman diagrams contributing to $S_\mathrm{hl}$ at N$^3$LO.
Time-like Wilson lines, light-like Wilson lines, gluons, and light fermions are
 represented by solid double lines, dashed double lines, curly lines, and solid lines, respectively. The direction of the Wilson lines and the prescription for the routing of the external momentum $\omega$ are the same as in the one-loop diagram on the right of \fig{cutIm}.
\label{fig:diags}}
\end{figure*}

Our three-loop calculation of the soft function $S_\mathrm{hl}$ is based on the 
definition in \eq{dispersiveWL} and performed very much along the lines of our 
jet function calculation in \rcite{Bruser:2018rad}, to which we refer for 
more details.
We work in general covariant gauge with gauge parameter $\xi$, where $\xi=0$ 
corresponds to Feynman gauge. Ultraviolet (UV) and (intermediate) IR divergences
are regulated with dimensional regularization ($d=4-2\epsilon$).

We use $\texttt{qgraf}$~\cite{Nogueira:1991ex} to generate all relevant 
three-loop (propagator-type) Feynman graphs with one internal light-like and 
two external time-like Wilson lines (one incoming, one outgoing), like the 
ones in \fig{diags}.%
\footnote{
Although the relation of $S_\mathrm{hl}$ to the time-ordered product of Wilson 
lines in \eq{dispersiveWL} was not made explicit, also the NNLO computation of 
$S_\mathrm{hl}$ in \rcite{Becher:2005pd} was performed in terms of the same 
type of loop diagrams.}
The diagrams are further processed by an in-house $\texttt{Mathematica}$ code 
which assigns the corresponding Feynman rules and performs the necessary Dirac, 
Lorentz and color algebra.
After that the diagrams are given by linear combinations of scalar Feynman 
integrals.
These integrals can then be mapped onto 16 integral topologies with twelve 
linearly independent linear and quadratic propagators. 
The associated 16 integral families 
contain integrals with integer propagator 
powers ranging from minus three to plus five.
The mapping of Feynman integrals onto the different topologies requires
numerous multivariate partial fraction operations on products of linear Wilson 
line propagators followed by suitable shifts of the loop momenta.
In order to automatize the extensive partial fractioning we implemented the 
algorithm outlined in \rcite{Pak:2011xt} in our code.

Next, we perform the integration-by-parts (IBP) 
reduction~\cite{Chetyrkin:1981qh} of the integrals 
in each of the 16 families to a set of master integrals (MIs) using the public 
program $\texttt{FIRE5}$~\cite{Smirnov:2014hma}.%
\footnote{The plain IBP reduction with $\texttt{FIRE5}$ yields an overcomplete 
set of MIs. To obtain a minimal MI basis for each family we employ the 
algorithm of \rcite{Pak:2011xt} to identify equal Feynman integrals.
This algorithm is implemented in the $\texttt{FindRules}$ 
command of $\texttt{FIRE5}$, which we apply to a large list of test integrals 
in each family. The output are identities among these integrals, which must 
also hold after IBP reduction. Demanding this 
yields another eight independent 
relations between MIs belonging to the same family, see also 
\rcite{Bruser:2018rad}.}
We then identify pairs of equal MIs of different families that are related by 
shifts of their loop momenta. The resulting total set of MIs across the 16 
families still turns out to be redundant.
We find 14 additional relations involving at least three MIs of different 
families due to partial fraction identities among their linear propagators.
Finally, the three-loop contribution to the matrix element in 
\eq{dispersiveWL} can be expressed as a linear combination of 45 MIs belonging 
to nine different integral families.%
\footnote{The total number of linearly independent MIs across all families is 
64, but only 45 contribute to $S_\mathrm{hl}$.}
At this point we already notice that the gauge  
parameter $\xi$ manifestly cancels out in the sum of all diagrams indicating 
the correctness of our setup.
The 45 contributing MIs can be cast into the form
\begin{align}
G(\vec a, \vec b,\vec c\,) = \bigl(i\pi^{\frac{d}{2}}\bigr)^{-3}
\!\int\!\!
\frac{
\df^dk_1\,\df^dk_2\,\df^dk_3}{
\cD_1^{a_1}
\cD_2^{a_2}
\cD_3^{a_3}
\cD_4^{a_4}
\cD_5^{ a_5 }
\cD_6^{a_6}
\cD_7^{b_1}
\cD_8^{b_2}
\cD_9^{b_3}
\cD_{10}^{b_4}
\cD_{11}^{b_5 }
\cD_{12}^{c_1}
\cD_{13}^{c_2}
\cD_{14}^{c_3}
\cD_{15}^{c_4}
\cD_{16}^{c_5}}
\label{eq:Gintegral}
\end{align}
with the following (linearly-dependent) propagator denominators
\begin{align}
&\cD_1=-k_1^2\,, 
& &\!\!\cD_2=-k_2^2\,,
& &\!\!\cD_3=-k_3^2\,, 
& &\!\!\cD_4=-(k_1-k_2)^2\,,& \nn\\
&\cD_5=-(k_2-k_3)^2\,, 
& &\!\!\cD_6=-(k_3-k_1)^2\,,
& &\!\!\cD_7=-v\!\cdot\! k_1\,, 
& &\!\!\cD_8=-v\!\cdot\! k_2\,, & \nn\\
&\cD_9=-v\!\cdot\!k_3\,, 
& &\!\!\cD_{10}=-v\!\cdot\! (k_1-k_3)\,,
& &\!\!\cD_{11}=-v\!\cdot\! (k_2-k_3)\,,
& &\!\!\cD_{12}=-n\!\cdot\!k_1-\omega\,, & \nn\\
&\cD_{13} =-n\!\cdot\! k_2-\omega\,,
& &\!\!\cD_{14}=-n\!\cdot\! k_3-\omega\,,
& &\!\!\cD_{15}=-n\!\cdot\! (k_1-k_3)-\omega\,,
& &\!\!\cD_{16}=-n\!\cdot\! 
(k_2-k_3)-\omega\,,&
\end{align}
where the usual (causal) `$-i0$' prescription, i.e.\ $\cD_i \to \cD_i 
-i0$, is understood.
The nine integral families containing the 45 MIs are defined by their maximal 
topologies with twelve linearly independent $\cD_i$. These topologies are 
determined by restricting the propagator powers in \eq{Gintegral}, for instance 
by
\begin{align}
&\text{topology 1:}\quad  b_4, b_5, c_4, c_5=0\,,
&\text{topology 2:}\quad  b_3, b_5, c_4, c_5=0\,,&\nn\\
&\text{topology 3:}\quad b_4, b_5, c_3, c_4=0\,,
&\text{topology 4:}\quad b_3, b_4, c_3, c_4=0\,,&\nn\\
&\text{topology 5:}\quad b_2, b_3, c_4, c_5=0\,,
&\text{topology 6:}\quad b_3, b_4, c_2, c_4=0\,,&\nn\\
&\text{topology 7:}\quad b_4, b_5, c_2, c_4=0\,,
&\text{topology 8:}\quad b_3, b_5, c_3, c_4=0\,,&\nn\\
&\text{topology 9:}\quad b_3, b_4, c_1, c_5=0\,.&
\end{align}

From the scaling properties of the integrand in \eq{Gintegral} for general 
time-like vector $v^\mu$ and light-like vector $n^\mu$ we conclude
\begin{align}
G(\vec a, \vec b,\vec c \,)  =
\big(v^2\big)^{\frac{3}{2}d-A-B}
(n\!\cdot\! v)^{2A +B-3d}
(-\omega-i0)^{3d-2A -B-C} 
I(\vec a, \vec b,\vec c,\epsilon)
\label{eq:Ganswer}
\end{align}
with $A=\sum_i a_i$, $B=\sum_i b_i$, and $C=\sum_i c_i$.
The dependence on the external kinematics thus totally
factors out and we are left to compute the dimensionless function
$I(\vec a, \vec b,\vec c,\epsilon)$ as an expansion in $\eps$. %
For the case of heavy-to-light decays calculated in the rest frame of the heavy 
quark we have $v^2=n\!\cdot\!v=1$ by definition.
For convenience we set $\omega=-1$ during the calculation of the MIs and 
restore their $\omega$ dependence later.
Twelve MIs are simple enough to be evaluated by direct integrations over the 
associated Feynman parameters in $d=4-2\epsilon$ dimensions. 
The results involve hypergeometric and gamma functions and are expanded in 
$\eps$ with the help of the $\texttt{Mathematica}$ package 
$\texttt{HypExp2}$~\cite{Huber:2007dx}. 

To solve the remaining 33 MIs we proceed in the same way as for our jet 
function calculation in \rcite{Bruser:2018rad}. The method was inspired by 
\rcites{Panzer:2014gra,vonManteuffel:2014qoa,vonManteuffel:2015gxa}.
The key idea is to express the 33 MIs as a linear combination of quasi-finite 
integrals and known MIs. 
Quasi-finite integrals are free of (endpoint) divergences from the 
integrations in the Feynman parameter representation (at the Euclidean point 
$\omega=-1$) for some (even) integer dimension. 
Starting from a given MI in $d=4-2\eps$ one can construct a corresponding 
quasi-finite integral by raising the spacetime dimension by an even number 
and/or increasing appropriate propagator powers by integer amounts.
The former decreases (increases) the degree of IR (UV) divergence, whereas the 
latter decreases (but not necessarily increases) the degree of UV (IR) 
divergence.
To systematically identify suitable quasi-finite integrals 
we employ the dedicated algorithm implemented in the public program 
$\texttt{Reduze2}$~\cite{vonManteuffel:2012np}.
For our purposes we find 18 integrals that are 
quasi-finite in $4-2\epsilon$ and 15 integrals that are quasi-finite in 
$6-2\epsilon$ dimensions.
To compute them in the respective dimension we first expand their nonsingular 
integrands in the Feynman parameter representation to high enough order in 
$\eps$. We then perform the integrations with the help of 
$\texttt{HyperInt}$~\cite{Panzer:2014caa}, a powerful computer algebra package 
for the analytical evaluation of convergent linearly reducible (Feynman) 
integrals in terms of multiple polylogarithms. 
The quasi-finite integrals (in their respective dimension) are related to the 
original MIs (in $d=4-2\eps$) by dimensional 
recurrence~\cite{Tarasov:1996br,Lee:2009dh,Lee:2010wea} and IBP reduction.
To determine the relevant dimensional recurrence relations between integrals 
in $d$ and $d+2$ dimensions we use the public code 
$\texttt{LiteRed}$~\cite{Lee:2012cn,Lee:2013mka}.
Our choice of the 33 quasi-finite integrals is such that their results together 
with the 12 already computed MIs uniquely determine the remaining 33 MIs.
We successfully verified all analytic expressions for the MIs obtained in 
this way numerically using the sector decomposition program 
$\texttt{FIESTA4}$~\cite{Smirnov:2015mct}.
Finally we insert the results for the 45 MIs in the IBP reduced expression for 
each three-loop Feynman diagram contributing to $S_\mathrm{hl}$ and expand to 
the required order in $\eps$, see below. 
We also repeated the calculation for the relevant lower-order graphs using the 
same setup.

%%%%%%%%%%%%%%%%%%%%%%%%%%%%%%%%%%%%%%%%%%%%%%%%%%%%%%%%%%%%%%%%%%%%%%%%%%%%%%%%
\section{Results}
%%%%%%%%%%%%%%%%%%%%%%%%%%%%%%%%%%%%%%%%%%%%%%%%%%%%%%%%%%%%%%%%%%%%%%%%%%%%%%%%
\label{sec:res}

After computing the relevant Feynman diagrams as described in the 
previous section we take their imaginary part according to \eq{dispersiveWL} 
using
\begin{align}
\mathrm{Im} \bigg[ (-\omega-i0)^{-1-a \eps}  \bigg]  &= 
 - \sin (\pi a \eps)  \, \theta(\omega)\,  \omega^{-1-a \eps}
 \,.
\end{align}
Adding the contributions of all diagrams (including the lower-order ones) we 
obtain the bare soft function%
\footnote{Here we consistently set $v^2=n\!\cdot\!v=1$. If needed, the 
dependence on the scalar products $v^2$ and $n\!\cdot\!v$ can be reconstructed 
straightforwardly using the scaling properties of the matrix element in 
\eq{Shldef}, cf.\ \eq{Ganswer}.}
\begin{align}
S^\bare_\mathrm{hl}(\omega) ={}&
1+\frac{\alpha_s^\bare}{4\pi}\, \theta(\omega)\, \omega^{-1-2\epsilon}
\,C_F \,K_F \nn\\
&+\bigg(\frac{\alpha_s^\bare}{4\pi}\bigg)^{\!2}
\theta(\omega)\,\omega^{-1-4\epsilon}\,
\bigg(C_F^2 \,K_{FF}+ C_FC_A \,K_{FA}+ C_F n_f T_F \, K_{F f}\bigg) 
\nn\\
&+\bigg(\frac{\alpha_s^\bare}{4\pi}\bigg)^{\!3}
\theta(\omega)\,\omega^{-1-6\epsilon}\,
\bigg(C_F^3 \,K_{FFF}+ C_F^2C_A \,K_{FFA}+ C_F C_A^2 \,K_{FAA} 
 \nn\\
&\qquad + C_F^2 n_f T_F\, K_{FFf} + C_F C_A n_f T_F \, K_{F A f}
+C_F (n_f T_F)^2 \,K_{Fff}\bigg) +\ord{\alpha_s^4} 
\label{eq:baresoftfunc}
\end{align}
in terms of the bare coupling $\alpha_s^\bare = Z_\alpha \mu^{2 
\eps}\alpha_s$ with $n_f$ being the number of light (massless) quark flavours. 
The color constants of the $SU(N_c)$ gauge group are $C_A=N_c$, 
$C_F=(N_c^2-1)/(2N_c)$, and $T_F=1/2$.  
The coefficients $K_X$ of each color structure are given in \app{baredata}.
For illustration we show in \fig{diags} for each of the six three-loop $K_X$ 
coefficients one sample Feynman diagram (arranged in the corresponding order) 
that contributes to it.
Throughout this work we employ the $\msb$ renormalization scheme.
The relevant terms of the strong coupling renormalization factor 
$Z_\alpha$ are
\begin{align}
 Z_\alpha=1+\frac{\alpha_s}{4\pi}\left(-\frac{\beta_0}{\epsilon}\right)
 +\left(\frac{\alpha_s}{4\pi}\right)^2\left(\frac{\beta_0^2}{\epsilon^2}
 -\frac{\beta_1}{2\epsilon}\right)
 +{\cal O}\left(\alpha_s^3\right)
\label{eq:Zalpha}
\end{align}
with
\begin{align}
 \beta_0= \frac{11}{3}C_A-\frac{4}{3}T_Fn_f  \,,
 \qquad 
 \beta_1= \frac{34}{3}C_A^2-\frac{20}{3}C_A n_f T_F - 4 C_F n_f T_F \,.
\end{align}
For the $\eps$ expansion of \eq{baresoftfunc} we employ the 
distributional identity
\begin{align}
 \mu^{a \eps}\,\theta(\omega)\, \omega^{-1-a \eps} &= - 
\frac{\delta(\omega)}{a \eps}  + 
\sum_{n=0}^{\infty} \frac{(-a\eps)^n}{n!} 
\frac{1}{\mu} \cL_{n}\bigg(\frac{\omega}{\mu}\bigg)
\label{eq:distID}
\end{align}
with the usual plus distributions defined as
%%%
\begin{align} \label{eq:cLn}
\cL_n(x)
&= \biggl[ \frac{\theta(x) \ln^n \!x}{x}\biggr]_+ \!\!
= \lim_{\eps \to 0} \frac{\df}{\df x}\biggl[ \theta(x- \eps)\frac{\ln^{n+1} 
x}{n+1} \biggr]\,.
\end{align}
The bare and renormalized soft functions are related by
\begin{align}
\label{eq:barerenrel}
S_\mathrm{hl}^\bare(\omega)  &=  Z^S(\omega,\mu) \otimes 
S_\mathrm{hl}(\omega,\mu)
\,,
\end{align}
where the $\otimes$~symbol denotes a convolution of the type
\begin{align}
A(\omega) \otimes B(\omega) \equiv \int\! \df \omega'\, 
A^i(\omega-\omega')\, B(\omega') \,.
\label{eq:conv}
\end{align}
Convolutions among the plus distributions $\cL_n$ take the form
%%%
\begin{align} \label{eq:ExpLnLm}
\cL_m(\omega) \otimes \cL_n(\omega)
= V_{-1}^{mn}\, \delta(\omega) + \sum_{k=0}^{m+n+1} V_k^{mn}\, \cL_k(\omega)\,.
\end{align}
%%%
A generic expression for $V_k^{mn}$ can be found in \rcite{Ligeti:2008ac}.

%%%%%%%%%%%%%%%%%%%%%%%%%%%%%%%%%%%%%%%%%%%%%%%%%%%%%%%%%%%%%%%%%%%%%%%%%%%%%%%%
\subsection{Anomalous Dimension}
\label{subsec:anomdim}
%%%%%%%%%%%%%%%%%%%%%%%%%%%%%%%%%%%%%%%%%%%%%%%%%%%%%%%%%%%%%%%%%%%%%%%%%%%%%%%%

The RGE of our 1-jettiness soft function reads
\begin{align}
\label{eq:RGE}
\mu \frac{\df}{\df \mu} S_\mathrm{hl}(\omega, \mu) &= 
\Gamma^S(\omega,\mu) \otimes S_\mathrm{hl}(\omega, \mu) \,,
\end{align}
with the anomalous dimension
%%%
\begin{align} 
\Gamma^S(\omega, \mu)
&= - \big[Z^S(\omega, \mu)\big]^{-1} 
\otimes \mu \frac{\df}{\df \mu} Z^S(\omega,\mu)
\label{eq:Zgammarel}
\\
&= 2 \Gamma^q_{\cusp}(\as)\,\frac{1}{\mu}\cL_0\bigg(\frac{\omega}{\mu}
\bigg) + \gamma^S(\as)\,\delta(\omega)
\label{eq:anomdim}
\,.\end{align}
For the loop expansion of the anomalous dimensions we adopt the notation
\begin{align}
 \Gamma^q_\mathrm{cusp}(\alpha_s)&=\sum_{n=0}^{\infty} \Gamma^q_n
 \left(\frac{\alpha_s}{4\pi}\right)^{\!n+1}\,,
% \qquad  \Gamma^q_n = C_F \Gamma_n \quad  \text{for } n=0,1,2\,, \\
\qquad \gamma^S(\alpha_s) =\sum_{n=0}^{\infty}\gamma^S_n
\left(\frac{\alpha_s}{4\pi}\right)^{\!n+1}\,.
\label{eq:anomdimexp}
\end{align}
With the soft renormalization factor $Z^S$ determined from our bare results in 
\eq{Zgammarel} we obtain
\begin{align}
\gamma^S_0 ={}& 4 C_F\,,  \label{eq:gammaS0} \\
\gamma^S_1 ={}& C_F \bigg[
C_A \bigg(36 \zeta_3-\frac{220}{27}-\frac{\pi ^2}{9}\bigg)
- n_f T_F \bigg(\frac{16}{27}+\frac{4 \pi ^2}{9}\bigg)\bigg]\,, 
\label{eq:gammaS1}\\
\gamma^S_2 ={}& 
\label{eq:gammaS2}
C_F \bigg[
C_A^2 \bigg(\frac{5428 \zeta_3}{9}-\frac{64 \pi ^2 \zeta_3}{9}-264 \zeta_5
-\frac{81215}{729}+\frac{853 \pi ^2}{243}-\frac{44 \pi ^4}{45}\bigg) \\
&\quad+C_A n_f T_F \bigg(-\frac{4432 \zeta_3}{27}+\frac{4460}{729}
-\frac{1388 \pi ^2}{243}+\frac{16 \pi ^4}{15}\bigg) \nn\\
&\quad+C_F n_f T_F \bigg(-\frac{32 \zeta_3}{9}+\frac{1442}{27}
-\frac{4 \pi ^2}{3}-\frac{16 \pi ^4}{45}\bigg) 
%\nn\\
%&\quad\qquad
+(n_f T_F)^2 \bigg(-\frac{448 \zeta_3}{27}+\frac{6592}{729}
+\frac{80 \pi ^2}{81}\bigg)\bigg]\,, \nn
\end{align}
in addition to the known terms of the cusp anomalous dimension given in 
\app{anomdims}.
The one- and two-loop results in \eqs{gammaS0}{gammaS1} agree with those in 
\rcite{Becher:2005pd} (after adapting to their conventions).
In the following we relate the soft anomalous dimension to corresponding 
collinear and hard anomalous dimensions in SCET factorization in order to 
verify our results.
As we will see $\gamma^S_2$ can thus also be determined indirectly, 
i.e.\ without a dedicated three-loop calculation, using 
known results. However, to the best of our knowledge, the explicit expression 
in \eq{gammaS2} has not been given in the literature so far.

The anomalous dimension associated with the virtual IR singularities due to 
strong interactions among onshell partons in a squared QCD scattering amplitude 
can be understood as the anomalous dimension of a corresponding hard function 
in SCET.
It is therefore intrinsically tied to the UV divergences of soft and collinear 
operator matrix elements in SCET by RG consistency.
The generic all-order structure of the anomalous dimension for QCD amplitudes 
involving massive quarks was derived in \rcite{Becher:2009kw}.
For the heavy-to-light decay with one massless and one massive external quark 
the it is given by%
\footnote{Here and in the following we suppress a $+i0$ accompanying the scalar 
product in the argument of the logarithm, which is necessary for the analytic 
continuation to other kinematical situations, where e.g.\ both quarks are 
outgoing/incoming.}
\begin{align}
 \Gamma_\mathrm{hl} = -\Gamma^q_\mathrm{cusp}(\alpha_s)\ln\frac{\mu}{2v\cdot p}
 +\gamma^q+\gamma^Q\,,
 \label{eq:Gammahl}
\end{align}
where $p$ is the (outgoing) four-momentum of the massless quark ($p^2=0$, 
$v\cdot p = m_b/2$) and 
$\Gamma^q_\mathrm{cusp}(\alpha_s)$ is the light-like cusp anomalous dimension 
in the fundamental representation of $SU(N_c)$. 
The noncusp anomalous dimensions $\gamma^q$ and $\gamma^Q$ are associated with 
each massless and massive external quark, respectively.
They accordingly contribute to the anomalous dimension of any QCD scattering 
amplitude with multiple quark legs~\cite{Becher:2009kw} and are in that sense 
universal.
The renormalized SCET hard function $H_\mathrm{hl}$ corresponds to the finite 
part of the respective QCD amplitude squared, i.e.\ where all IR and UV 
divergences have been subtracted.
We thus have
\begin{align}
\mu \frac{\df}{\df \mu}H_\mathrm{hl} = 2\Gamma_\mathrm{hl} H_\mathrm{hl}\,.
\end{align}
For the QCD amplitude with two external heavy quarks (one outgoing, one 
incoming, where $p_i^2=m_i^2$, $p_1\!\cdot p_2>0$) the (hard) anomalous 
dimension reads
\begin{align}
 \Gamma_\mathrm{hh} = \Gamma^Q_\mathrm{cusp}(\beta,\alpha_s) + 2\gamma^Q\,.
 \label{eq:Gammahh}
\end{align}
Here the angle-dependent cusp anomalous dimension 
$\Gamma^Q_\mathrm{cusp}(\beta,\alpha_s)$
with (Minkowskian) cusp angle $\beta = \mathrm{arccosh}(\frac{p_1 \!\cdot 
p_2}{m_1 m_2})$ is defined such that in the large angle expansion,%
\footnote{Note that in the literature traditionally often the full 
$\Gamma_\mathrm{hh}$ 
is referred to as the angle-dependent cusp anomalous dimension, see e.g.\ 
\rcites{Korchemsky:1991zp,Grozin:2015kna}.}
\begin{align}
 \Gamma^Q_\mathrm{cusp}(\beta,\alpha_s) = 
 \Gamma^q_\mathrm{cusp}(\alpha_s) \, \beta + \ORD{\frac{1}{\beta}}
 \,,
 \label{eq:cusplargebeta}
\end{align}
there is no $\ord{\beta^0}$ term.
As the large angle limit corresponds to the 
limit where the mass of one or both of the quarks vanishes it is not surprising 
that the coefficient of the leading term in \eq{cusplargebeta} equals the 
light-like cusp anomalous dimension~\cite{Korchemsky:1985xj,Korchemsky:1991zp}.

For completeness and comparison we also recall the corresponding anomalous 
dimension for a QCD amplitude with two massless quarks ($p_i^2=0$, $p_1\!\cdot 
p_2>0$):
\begin{align}
 \Gamma_\mathrm{ll} = -\Gamma^q_\mathrm{cusp}(\alpha_s) 
 \ln\frac{\mu^2}{2 p_1\!\cdot p_2} + 2\gamma^q\,.
 \label{eq:Gammall}
\end{align}
We stress that $\Gamma^q_\mathrm{cusp}(\alpha_s)$ and $\gamma^q$ are the 
same as in \eq{Gammahl}.

Renormalization group invariance of the decay 
rate in \eq{fact} requires
\begin{align}
 \mu \frac{\df}{\df \mu}\left(H_\mathrm{hl}\times S_\mathrm{hl} 
 \otimes J_q\right)=0\,.
\end{align}
For the noncusp anomalous dimensions this implies%
\footnote{In our convention the jet function RGE is analogous to \eq{RGE}.}
\begin{align}
 2\gamma^q+2\gamma^Q + \gamma^S +\gamma^{J_q} = 0\,.
 \label{eq:RGgammarel}
\end{align}
The three-loop contribution $\gamma^q_2$ was obtained from the calculation of 
the three-loop massless quark form factor~\cite{Moch:2005id} via
\eq{Gammall}. In 
\rcite{Bruser:2018rad} we directly computed the massless quark jet function 
anomalous dimension $\gamma_2^{J_q}$.
It was initially derived indirectly from the RG invariance of the factorized 
DIS cross section in the threshold region~\cite{Becher:2006mr} using the 
three-loop results of \rcites{Moch:2005id,Moch:2004pa}. 
The heavy quark noncusp anomalous dimension $\gamma^Q_2$ can be extracted from 
the three-loop result of $\Gamma_\mathrm{hh}$ in \rcite{Grozin:2015kna} using 
\eqs{Gammahh}{cusplargebeta}.
In fact it can be read off directly from the (nonlogarithmic) constant in the 
large-angle expansion of $(-\Gamma_\mathrm{hh})$ explicitly performed in 
appendix B of \rcite{Hoang:2015vua}.
We have
\begin{align}
 \gamma^Q_2={}&
C_F \bigg[ C_A^2\bigg(-\frac{4}{3} \pi ^2 \zeta _3-\frac{740 \zeta_3}{9}
+36 \zeta _5-\frac{22 \pi ^4}{45}+\frac{304 \pi ^2}{27}-\frac{343}{9}\bigg) 
+ C_F n_f T_F\bigg(\frac{110}{3}-32\zeta _3\bigg)  \nn\\
&\quad + C_A n_f T_F \bigg(\frac{496 \zeta _3}{9}-\frac{80 \pi ^2}{27}
+\frac{356}{27}\bigg)
+\frac{32}{27} (n_f T_F)^2  \bigg]\,.
\label{eq:gammaQ2}
\end{align}
We give the explicit expressions for $\gamma^q_2$ and $\gamma_2^{J_q}$ in 
\app{anomdims}.
We can now solve \eq{RGgammarel} for $\gamma^S_2$ and find exact agreement with 
\eq{gammaS2}.
This serves as a valuable cross check of our three-loop calculation of 
$S_\mathrm{hl}$.
At the same time it confirms the prediction~\cite{Becher:2009kw} regarding the 
two-parton correlation part of the IR singularity structure of QCD scattering 
amplitudes with massive external quarks according to 
\eqs{Gammahl}{Gammahh}.

%%%%%%%%%%%%%%%%%%%%%%%%%%%%%%%%%%%%%%%%%%%%%%%%%%%%%%%%%%%%%%%%%%%%%%%%%%%%%%%%
\subsection{Renormalized results}
%%%%%%%%%%%%%%%%%%%%%%%%%%%%%%%%%%%%%%%%%%%%%%%%%%%%%%%%%%%%%%%%%%%%%%%%%%%%%%%%
\label{subsec:renres}

Upon $\msb$ renormalization the coefficients in the loop expansion of the 
1-jettiness soft function for heavy-to-light quark decays 
\begin{align}
S_\mathrm{hl}(\omega, \mu) = \sum_{m=0}^\infty 
\bigg(\frac{\alpha_s}{4\pi}\bigg)^{\!m}
S^{(m)}(\omega,\mu)
\end{align}
take the form
\begin{align}
S^{(m)}(\omega,\mu)=S^{(m)}_{-1}\delta(\omega)+\sum_{n=0}^{2m-1}S^{(m)}_{n}\frac
{1}{\mu}\cL_n\left(\frac{\omega}{\mu}\right)\,.
\end{align}
By iteratively solving the RGE in \eq{RGE} as an expansion in $\alpha_s$ the 
terms depending on the 
renormalization scale $\mu$, i.e.\ the coefficients $S^{(m)}_{n}$ with $n\ge0$, 
are completely determined by the lower-order constants $S^{(l<m)}_{-1}$ and 
anomalous dimension coefficients.
To three-loop order we have
\begin{align}
S^{(1)}_{1} ={}& -2 \Gamma^q_0  \,, \label{eq:S11}
\\
S^{(1)}_{0} ={}& - \gamma _0^S\,,
\\
S^{(2)}_{3} ={}& 2 \big(\Gamma^q_0\big)^2\,,
\\
S^{(2)}_{2} ={}& \Gamma^q _0 \Big(2\beta _0+3 \gamma _0^S\Big)\,,
\\
S^{(2)}_{1} ={}& -\frac{2\pi ^2}{3}\big(\Gamma^q_0\big)^2-2\Gamma^q _1+ 
\gamma_0^{S}\Big(\gamma_0^{S}+2 \beta_0 \Big) -2S^{(1)}_{-1} \Gamma^q_0 
\,,
\\
S^{(2)}_{0} ={}& 4\big(\Gamma^q_0\big)^2 \zeta _3-\frac{\pi ^2 }{3} \Gamma^q _0 
\gamma_0^S- 
\gamma _1^S
-S^{(1)}_{-1} \Big(2\beta _0+\gamma _0^S\Big)
\,,
\\
S^{(3)}_{5} ={}& -\big(\Gamma^q_0\big)^3\,,
\\
S^{(3)}_{4} ={}& -\big(\Gamma^q_0\big)^2 \biggl(\frac{5\gamma 
_0^S}{2}+\frac{10}{3}\beta_0\biggr)\,,
\\
S^{(3)}_{3} ={}& \frac{4  \pi ^2}{3} \big(\Gamma^q_0\big)^3
+4\Gamma^q_0 \Gamma^q _1- \frac{8}{3} \Gamma^q _0  \beta _0^2
-2\Gamma^q _0 \gamma _0^{S}\bigg(\frac{10}{3} \beta _0 + \gamma _0^{S}\bigg)
+2 S^{(1)}_{-1} \big(\Gamma^q_0\big)^2\,,
\\
S^{(3)}_{2} ={}& -20\zeta_3 \big(\Gamma^q_0\big)^3
+2\pi^2 \big(\Gamma^q_0\big)^2\Big(\beta _0 +\gamma _0^S\Big)
+\Gamma^q _0\Big(2\beta _1+3 \gamma _1^S\Big)
+\Gamma^q _1\Big(4 \beta _0+3\gamma _0^S\Big)
-\frac{1}{2}\big(\gamma _0^{S}\big)^3 \nn\\
&-\beta _0 \gamma _0^S\Big(4\beta _0+3\gamma _0^{S}\Big)
+S^{(1)}_{-1} \Gamma^q _0 \Big(8\beta _0+3\gamma _0^S\Big)\,,
\\
S^{(3)}_{1} ={}& \frac{2\pi^4}{45} \big(\Gamma^q_0\big)^3 
-8\zeta_3\big(\Gamma^q_0\big)^2\Big(2\gamma_0^S+3\beta_0 \Big) 
+2\pi ^2 \Gamma^q _0 \gamma _0^S\bigg(\frac{\gamma _0^{S} }{3} +\beta_0\bigg)
-\frac{4 \pi ^2}{3} \Gamma^q _0 \Gamma^q _1 
-2 \Gamma^q_2+2 \beta _1 \gamma _0^S
\nn\\
&+2\gamma _1^S \Bigl(\gamma _0^S +2 \beta _0\Bigr)
+S^{(1)}_{-1} \biggl[8 \beta_0^2 -\frac{2\pi ^2}{3} \big(\Gamma^q_0\big)^2
-2\Gamma^q _1 +\gamma _0^S\Big(\gamma_0^{S}+6\beta _0\Big) \biggr]
-2 \Gamma^q _0 S^{(2)}_{-1}\,,
\\
S^{(3)}_{0} ={}& \biggl(\frac{8}{3} \pi ^2 \zeta_3 - 24 \zeta _5\biggr) 
\big(\Gamma^q_0\big)^3 
+\frac{\pi^4 }{45}\big(\Gamma^q_0\big)^2\Big(4\beta _0+\gamma _0^S\Big)
-2\zeta _3 \Gamma^q _0\gamma _0^{S}\Big(2\beta_0+\gamma _0^{S}\Big)
+8\zeta _3\Gamma^q _0 \Gamma^q _1 -\gamma_2^S
\nn\\
&-\frac{\pi^2 }{3} \Big(\Gamma^q _1 \gamma_0^S+\Gamma^q _0 \gamma _1^S\Big) 
+S^{(1)}_{-1}\biggl[4 \zeta_3\big(\Gamma^q_0\big)^2-\frac{\pi ^2}{3}\Gamma^q _0 
\big(\gamma _0^S+2\beta _0\big) -\gamma_1^S-2 \beta _1\biggr]
\nn\\
&-S^{(2)}_{-1}\Bigl( \gamma_0^S+4\beta _0\Bigr)\,.
\label{eq:S03}
\end{align}
Our explicit calculation of $S_\mathrm{hl}(\omega,\mu)$ perfectly reproduces 
\eqsto{S11}{S03}, which serves as a cross check.
In addition it yields the delta function coefficients
\begin{align}
S^{(1)}_{-1}={}& -\frac{\pi ^2}{6}  C_F \,,
\nn\\
S^{(2)}_{-1}={}&  C_F^2\biggl(32 \zeta _3-\frac{3 \pi ^4}{40}
-\frac{4\pi^2}{3}\biggr) 
+C_A C_F\biggl(-\frac{107 \zeta _3}{9}+\frac{67 \pi^4}{180}
-\frac{427 \pi^2}{108}-\frac{326}{81}\biggr) \nn\\
&+C_F n_f T_F\biggl(-\frac{20 \zeta _3}{9}+\frac{5 
\pi^2}{27}-\frac{8}{81}\biggr) 
\,,\\
S_{-1}^{(3)}={}& C_F^3\biggl(
-\frac{1280 \zeta _3^2}{3}+80 \pi ^2 \zeta_3-\frac{64 \zeta _3}{3}
-768\zeta_5 + \frac{3097 \pi ^6}{9072}+\frac{26 \pi ^4}{45}\biggr) \nn\\
&+C_F^2 C_A 
\biggl(288 \zeta _3^2+\frac{1883 \pi ^2 \zeta _3}{54}+\frac{1504\zeta _3}{27}
-\frac{2816 \zeta _5}{3}-\frac{\pi ^6}{360}+\frac{11287 \pi ^4}{3240}
+\frac{1483\pi ^2}{243}\biggr) \nn\\
&+ C_F C_A^2\Big(
-\frac{1052\zeta _3^2}{9}-\frac{136 \pi ^2 \zeta_3}{3}
+\frac{998 \zeta_3}{243}+\frac{4369 \zeta _5}{9}
-\frac{13387 \pi^6}{51030}+\frac{6223 \pi^4}{972} \nn\\
&\qquad\qquad+\frac{45139 \pi ^2}{8748}-\frac{2662195}{26244}\Big) \nn\\
&+C_F^2 n_f T_F \biggl(-\frac{782}{27} \pi ^2 \zeta _3-\frac{3224\zeta_3}{81}
+\frac{2848 \zeta _5}{9}-\frac{673 \pi ^4}{810}
+\frac{695 \pi^2}{486}+\frac{11929}{486}\biggr) \nn\\
&+C_F C_A  n_f T_F 
\biggl(\frac{44 \pi ^2 \zeta _3}{3}
+\frac{464 \zeta _3}{81}-104 \zeta _5
-\frac{1169 \pi ^4}{1215}+\frac{121\pi^2}{2187}
+\frac{131659}{6561}\biggr) \nn\\
& +C_F (n_f T_F)^2 
 \biggl(\frac{736 \zeta _3}{243}-\frac{52 \pi ^4}{1215}+\frac{8 \pi^2}{243}
 +\frac{33920}{6561}\biggr) \,.
 \label{eq:S3const}
\end{align}
The expression for $S_{-1}^{(3)}$ is new and represents together with the 
three-loop soft anomalous dimension in \eq{gammaS2} the main result of this 
work.

%%%%%%%%%%%%%%%%%%%%%%%%%%%%%%%%%%%%%%%%%%%%%%%%%%%%%%%%%%%%%%%%%%%%%%%%%%%%%%%%
\section{Summary}
%%%%%%%%%%%%%%%%%%%%%%%%%%%%%%%%%%%%%%%%%%%%%%%%%%%%%%%%%%%%%%%%%%%%%%%%%%%%%%%%
\label{sec:sum}

In this paper we calculated the 1-jettiness ($\mathcal{T}_1$) soft function for 
heavy-to-light quark decays at N$^3$LO.
The renormalized result is given in \subsec{renres}.
The three-loop delta-function coefficient in \eq{S3const} and the three-loop 
contribution to the soft noncusp anomalous dimension in \eq{gammaS2} represent 
the genuinely new information at this order.
In \app{anomdims} we also collect all other noncusp anomalous dimensions 
required for N$^3$LL resummed heavy-to-light decay rates that are differential 
in either $\mathcal{T}_1$ or closely related observables like the photon energy 
in $B\to X_s \gamma$ or the jet invariant mass in $B\to X_u \ell \bar\nu$.
We explicitly checked the relation between hard, soft, and jet anomalous 
dimensions required by RG consistency. 
This also confirms the predicted universal structure~\cite{Becher:2009kw} of 
the IR singularities of QCD amplitudes due to two-parton interactions involving 
massive external quarks at three loops.
That is because we used this prediction to derive the three-loop 
hard anomalous dimension for heavy-to-light decays from the known three-loop IR 
singularities of the massive (heavy-heavy) and massless (light-light) quark form 
factors.

For N$^3$LL$^\prime$ accuracy also the three-loop contributions to the 
hard, jet, and soft functions in the corresponding $\mathcal{T}_1$-type 
factorization theorems for the decay rates are needed.
Our new soft function result represents together with the three-loop 
contribution to the jet function, which we computed in \rcite{Bruser:2018rad}, 
the two universal (i.e.\ process-independent) ingredients at this order. 
As such they also play a crucial role in the calculation of differential 
N$^3$LO heavy-to-light quark decay rates using the $N$-jettiness IR subtraction 
(slicing) method, e.g.\ for $t\to W^+(l^+\nu)b$.
The three-loop calculations of the corresponding process-dependent 
heavy-to-light hard functions are presumably feasible
using state-of-the-art multi-loop technology and may be performed in the not 
too far future.

%%%%%%%%%%%%%%%%%%%%%%%%%%%%%%%%%%%%%%%%%%%%%%%%%%%%%%%%%%%%%%%%%%%%%%%%%%%%%%%%
\begin{acknowledgments}
MS thanks John Collins, Iain Stewart, and Hua Xing Zhu for valuable 
discussions. ZLL thanks Yu-Ming Wang for helpful discussions.
This work was supported in part by the Deutsche Forschungsgemeinschaft through 
the project ``Infrared and threshold effects in QCD'', 
the Cluster of Excellence “Precision Physics, Fundamental Interactions, and 
Structure of Matter” (PRISMA$^+$ EXC 2118/1), and the 
Mainz Institute for Theoretical Physics (MITP) of PRISMA$^+$ (Project ID 
39083149).
RB was supported in part by the Deutsche Forschungsgemeinschaft under grant  
396021762 - TRR 257.
\end{acknowledgments}
%%%%%%%%%%%%%%%%%%%%%%%%%%%%%%%%%%%%%%%%%%%%%%%%%%%%%%%%%%%%%%%%%%%%%%%%%%%%%%%%
\appendix

\section{Hard, collinear, and cusp anomalous dimensions}
\label{app:anomdims}

For completeness we collect here the explicit expressions for all anomalous 
dimensions other than $\gamma^S$ in \eqsto{gammaS0}{gammaS2} relevant for 
the heavy-to-light quark decay up to three-loop order.
The convention for the loop expansion of the listed anomalous dimensions is 
analogous to \eq{anomdimexp}.

The one-, two-, and three-loop coefficients of the cusp anomalous dimensions 
are~\cite{Korchemsky:1987wg,Moch:2004pa}
\begin{align} 
\Gamma^q_0 ={}& 4C_F
\,,\\
\Gamma^q_1 ={}& 4C_F \bigg[\biggl( \frac{67}{9} -\frac{\pi^2}{3} \biggr)\,C_A  -
   \frac{20}{9}\,n_f T_F \bigg]
\,,\\
\Gamma^q_2 ={}& 4C_F \bigg[
\biggl(\frac{245}{6} -\frac{134 \pi^2}{27} + \frac{11 \pi ^4}{45}
  + \frac{22 \zeta_3}{3}\biggr)C_A^2
  + \biggl(- \frac{418}{27} + \frac{40 \pi^2}{27}  - \frac{56 \zeta_3}{3} 
\biggr)C_A n_f T_F
\nn\\
&\quad  + \biggl(- \frac{55}{3} + 16 \zeta_3 \biggr) C_F n_f T_F
  - \frac{16}{27}\, (n_f T_F)^2 \bigg]
\,.\end{align}
%%%
The four-loop coefficient $\Gamma^q_3$ (necessary for N$^3$LL resummation) is 
known completely numerically~\cite{Moch:2018wjh,Moch:2017uml}, while analytic 
expressions are at present available for all fermionic 
contributions~\cite{Grozin:2018vdn,Bruser:2019auj,Henn:2019rmi,Lee:2019zop}.

The hard noncusp anomalous dimension associated with massless 
external partons appearing in \eq{Gammahl} is given up to three loops
by~\cite{Idilbi:2006dg,Becher:2006mr}
\begin{align}
\gamma_0^q ={}& -3C_F \,,\\ 
\gamma_1^q={}&
C_F\bigg[C_A\biggl(26 \zeta_3-\frac{961}{54}-\frac{11\pi^2}{6}\biggr)
+C_F\biggl(-24 \zeta_3-\frac{3}{2}+2 \pi ^2\biggr)
+n_f T_F \biggl(\frac{130}{27}+\frac{2 \pi ^2}{3}\biggr) \bigg]
\,,\\ 
\gamma_2^q={}& 
C_F\bigg[C_A^2 \biggl(\frac{3526 \zeta_3}{9}-\frac{44 \pi ^2 \zeta_3}{9}
-136 \zeta_5 -\frac{139345}{2916}-\frac{7163 \pi ^2}{486}
-\frac{83 \pi^4}{90}\biggr)
\nn\\
&\quad +C_A C_F\biggl(-\frac{844 \zeta_3}{3}
-\frac{8 \pi ^2 \zeta_3}{3} - 120 \zeta_5 -\frac{151}{4}
+\frac{205 \pi^2}{9}+\frac{247 \pi ^4}{135}\biggr)
\nn\\
&\quad +C_F^2 \biggl(-68 \zeta_3+\frac{16 \pi ^2 \zeta_3}{3}+240 \zeta_5
-\frac{29}{2}-3 \pi^2-\frac{8 \pi^4}{5}\biggr)
\nn\\
&\quad +C_A n_f T_F \biggl(-\frac{1928 \zeta_3}{27}-\frac{17318}{729}
+\frac{2594 \pi ^2}{243}+\frac{22 \pi^4}{45}\biggr)
\nn\\
&\quad +C_F n_f T_F \biggl(\frac{512 \zeta_3}{9}+\frac{2953}{27}
-\frac{26\pi^2}{9}-\frac{28 \pi ^4}{27}\biggr)
\nn\\
&\quad +(n_f T_F)^2 \biggl(-\frac{32 \zeta_3}{27}+\frac{9668}{729}
-\frac{40 \pi^2}{27}\biggr)\bigg]
\,.
\end{align}

The hard noncusp anomalous dimension associated 
with the massive external quarks in \eqs{Gammahl}{Gammahh} has the coefficients
\begin{align}
\gamma_0^Q ={}& -2C_F \,, \label{eq:gammaQ0}\\ 
\gamma_1^Q={}& 
C_F \bigg[C_A\biggl(\frac{2\pi^2}{3}-\frac{98}{9}-4\zeta_3\biggr)
+n_f T_F \frac{40}{9}  \bigg] \,, \label{eq:gammaQ1}\\ 
\gamma_2^Q={}& 
C_F \bigg[ C_A^2\biggl(-\frac{4}{3} \pi ^2 \zeta _3-\frac{740 \zeta_3}{9}+36 
\zeta _5-\frac{22 \pi ^4}{45}+\frac{304 \pi ^2}{27}-\frac{343}{9}\biggr) 
+C_F n_f T_F \biggl(\frac{110}{3}-32\zeta _3\biggr)  \nn\\
&\quad+ C_A n_f T_F \biggl(\frac{496 \zeta _3}{9}-\frac{80 \pi 
^2}{27}+\frac{356}{27}\biggr)
+ \frac{32}{27}(n_f T_F)^2 \bigg]\,.
\end{align}
The one- and two-loop terms in \eqs{gammaQ0}{gammaQ1} can be found in 
\rcite{Becher:2009kw}. The three-loop contribution is copied for completeness 
from \eq{gammaQ2}.

The known terms of the noncusp quark jet function anomalous 
dimension are~\cite{Neubert:2004dd,Becher:2006mr}%
\footnote{Note that the $\gamma^q_n$ of 
\rcite{Bruser:2018rad} equal our $\gamma_n^{J_q}$.}
%%%
\begin{align}
\gamma_0^{J_q} ={}& 6 C_F \,,
\\
\gamma_1^{J_q} ={}&
C_F \bigg[C_A \biggl(-80 \zeta_3+\frac{1769}{27}+\frac{22 \pi 
^2}{9}\biggr)+C_F \biggl(48 \zeta_3+3-4 \pi 
^2\biggr)+ n_f T_F \biggl(-\frac{484}{27}-\frac{8 \pi ^2}{9}\biggr) \bigg] \,,
\\
\gamma_2^{J_q} ={}&
C_F \bigg[
C_A^2 \biggl(-\frac{11000 \zeta_3}{9}+\frac{176 \pi ^2 
\zeta_3}{9}+464 \zeta_5+\frac{412907}{1458}+\frac{838 \pi ^2}{243}+\frac{19 
\pi ^4}{5}\biggr)\
\nn\\
&\quad+C_A C_F \biggl(\frac{1688 \zeta_3}{3}+\frac{16 \pi ^2 
\zeta_3}{3}+240 \zeta_5+\frac{151}{2}-\frac{410 \pi ^2}{9}-\frac{494 \pi 
^4}{135}\biggr)
\nn\\
&\quad+C_F^2 \biggl(136 \zeta_3-\frac{32 \pi ^2 \zeta_3}{3}-480 \zeta_5+29+6 
\pi^2
+\frac{16 \pi ^4}{5}\biggr)
\nn\\
&\quad+ C_A n_f T_F \biggl(\frac{5312 
\zeta_3}{27}+\frac{10952}{729}-\frac{2360 \pi ^2}{243}-\frac{92 \pi
   ^4}{45}\biggr)
\nn\\  
&\quad+C_F n_f T_F 
\biggl(-\frac{416 \zeta_3}{9}-\frac{9328}{27}+\frac{64 \pi^2}{9}
+\frac{328 \pi^4}{135}\biggr)
\nn\\
&\quad+(n_f T_F)^2 \biggl(\frac{512 \zeta_3}{27}-\frac{27656}{729}
+\frac{160 \pi^2}{81}\biggr)\bigg]
\,.
\end{align}
%%%

\section{Bare data}
\label{app:baredata}

Here we present our expressions for the coefficients of the different color 
structures in the bare soft function, \eq{baresoftfunc}. We show the results as 
an expansion in $\eps = (4-d)/2$ to the order required for the calculation of 
the renormalized three-loop soft function using 
\eqss{Zalpha}{distID}{barerenrel}:
\begin{align}
K_F ={}&\frac{4}{\epsilon }-4+\frac{\pi^2 \epsilon }{3}
+\Bigg(-\frac{4 \zeta_3}{3}-\frac{\pi^2}{3}\Bigg) \epsilon^2
+\Bigg(\frac{4 \zeta_3}{3}+\frac{\pi ^4}{40}\Bigg)\epsilon^3 
   +\left(-\frac{1}{9} \pi ^2\zeta_3-\frac{4 \zeta_5}{5} 
   -\frac{\pi^4}{40}\right)\epsilon^4 \nn\\
   &+\left(\frac{2 \zeta_3^2}{9} +\frac{\pi ^2\zeta _3}{9}+\frac{4 \zeta _5}{5}
   +\frac{61 \pi^6}{30240}\right) \epsilon^5
   +\ord{\epsilon^6}
\,,\\ 
K_{FF} ={}& -\frac{8}{\epsilon ^3}+\frac{16}{\epsilon ^2}
+\frac{4 \pi ^2-8}{\epsilon }+\left(\frac{400 \zeta_3}{3}-8 \pi 
^2\right)+\left(-\frac{800 \zeta _3}{3}
+\frac{59 \pi ^4}{15}+4 \pi ^2\right)\epsilon \nn\\
   &+\left[\left(\frac{400}{3}-\frac{200 \pi ^2}{3}\right) \zeta _3
   +\frac{7696 \zeta _5}{5}-\frac{118 \pi^4}{15}\right]\epsilon^2\nn\\
   &+\left(-\frac{10000 \zeta _3^2}{9}+\frac{400 \pi ^2 \zeta_3}{3}
   -\frac{15392 \zeta _5}{5}+\frac{6229 \pi ^6}{1890}+\frac{59 \pi^4}{15}\right) 
\epsilon^3
+\ord{\epsilon ^4}
\,,\\
K_{FA} ={}& \frac{22}{3 \epsilon^2}
+\frac{1}{\epsilon}\left(\frac{2}{9}-\frac{2 \pi ^2}{3}\right)
    +\left(-36 \zeta_3+\frac{23\pi^2}{9}+\frac{220}{27}\right)
    +\Bigg(\frac{340 \zeta_3}{9}-\frac{67\pi^4}{45}+\frac{361\pi^2}{27}\nn\\
   &\quad +\frac{1304}{81}\Bigg) \epsilon 
   +\left[\left(\frac{404}{27}+\frac{334 \pi ^2}{9}\right) \zeta_3-492 \zeta _5
   +\frac{337 \pi ^4}{108}-\frac{4210 \pi ^2}{81}+\frac{7792}{243}\right] 
\epsilon^2\nn\\
   &+\Bigg[352 \zeta _3^2
   +\left(-\frac{23816}{81}-\frac{2158 \pi ^2}{27}\right) \zeta_3
   +\frac{13396 \zeta_5}{15}-\frac{6149 \pi^6}{3780}
   +\frac{10807\pi^4}{1620}\nn\\
   & \quad +\frac{36940\pi^2}{243}
   +\frac{46688}{729}\Bigg]\epsilon^3
   + \ord{\epsilon ^4}
\,,\\
K_{Ff} ={}& -\frac{8}{3 \epsilon ^2}+\frac{8}{9 \epsilon }
+\left(\frac{16}{27}-\frac{4 \pi^2}{9}\right)
+\left(\frac{112 \zeta_3}{9}+\frac{4 \pi ^2}{27}+\frac{32}{81}\right) \epsilon 
\nn\\
   &+\left(-\frac{112 \zeta _3}{27}-\frac{7 \pi ^4}{135}
   +\frac{8 \pi ^2}{81}+\frac{64}{243}\right)
   \epsilon ^2 \nn\\
   & +\left[\left(\frac{56 \pi ^2}{27}-\frac{224}{81}\right) \zeta _3
   +\frac{496 \zeta_5}{15}+\frac{7 \pi ^4}{405}
   +\frac{16 \pi ^2}{243}+\frac{128}{729}\right] \epsilon ^3
   +\ord{\epsilon^4}
\,,\\
K_{FFF} ={}& \frac{8}{\epsilon ^5}-\frac{24}{\epsilon ^4}
+\frac{24-14 \pi ^2}{\epsilon ^3}+\frac{-520 \zeta _3+42 \pi
   ^2-8}{\epsilon ^2}+\frac{1}{\epsilon}\left(1560 \zeta _3
   -\frac{185 \pi ^4}{12}-42 \pi ^2\right)\nn\\
   &+\Bigg[\left(910 \pi ^2-1560\right) \zeta _3-\frac{61464 \zeta _5}{5}
   +\frac{185 \pi ^4}{4}+14 \pi^2\Bigg]
   +\Bigg[16900 \zeta _3^2
   \nn\\
   &\quad +\left(520-2730 \pi ^2\right) \zeta _3+\frac{184392 \zeta_5}{5}
   -\frac{367153 \pi ^6}{15120}-\frac{185 \pi ^4}{4}\Bigg]\epsilon 
+\ord{\epsilon ^2}
\,, \\
K_{FFA} ={}& -\frac{22}{\epsilon ^4}
+\frac{1}{\epsilon ^3}\left(\frac{64}{3} +2 \pi ^2\right)  
+\frac{1}{\epsilon^2}\left(108 \zeta_3+\frac{107 \pi^2}{6}-\frac{214}{9}\right)
   +\frac{1}{\epsilon}\Bigg(842 \zeta _3\nn\\
   &\quad+\frac{59 \pi ^4}{30}-\frac{532 \pi^2}{9}-\frac{644}{27}\Bigg)
   +\Bigg[\left(-\frac{2888}{3}-343 \pi ^2\right) \zeta _3
   +\frac{31381 \pi^4}{720}\nn\\
   &\quad +1476 \zeta_5+\frac{12191 \pi ^2}{54}-\frac{3880}{81}\Bigg]
   \!+\!\Bigg[-6276 \zeta_3^2
   +\left(\frac{18686}{9}-\frac{1405 \pi^2}{6}\right)\zeta _3\nn\\
   &\quad+\frac{127086 \zeta_5}{5}-\frac{24313 \pi ^6}{5040}
   -\frac{1897 \pi ^4}{135}-\frac{47155 \pi^2}{81}-\frac{23312}{243}\Bigg] 
\epsilon +\ord{\epsilon ^2}
\,,\\
K_{FAA} ={}& \frac{484}{27 \epsilon ^3}
+\frac{1}{\epsilon^2}\left(\frac{2152}{81}-\frac{88 \pi ^2}{27}\right)
+\frac{1}{\epsilon}\left(-\frac{2288\zeta_3}{9}+\frac{44 \pi ^4}{135}
+\frac{619 \pi ^2}{81}+\frac{7414}{81}\right) \nn\\
   &+\Bigg[\left(\frac{64 \pi ^2}{9}
   -\frac{1204}{9}\right) \zeta _3+264 \zeta _5-\frac{77 \pi^4}{5}
   +\frac{39530\pi ^2}{243}+\frac{210311}{729}\Bigg]\nn\\
   & +\Bigg[\frac{2104 \zeta_3^2}{3} +\left(\frac{5144}{81}
   +\frac{6122 \pi ^2}{9}\right) \zeta _3-\frac{24974 \zeta_5}{3}+\frac{13387 
\pi ^6}{8505}-\frac{16517 \pi ^4}{3240}\nn\\
   &\quad-\frac{886981 \pi^2}{1458}+\frac{4205011}{4374}\Bigg] \epsilon 
   +\ord{\epsilon ^2}
\,,\\
K_{FFf} ={}& \frac{8}{\epsilon ^4}-\frac{32}{3 \epsilon ^3}
+\frac{1}{\epsilon^2}\left(-\frac{16}{9}-\frac{26\pi^2}{3}\right)
+\frac{1}{\epsilon }\left(-\frac{1208 \zeta_3}{3}+\frac{104 \pi^2}{9}
-\frac{428}{27}\right)\nn\\
   &\quad+\left(\frac{5120\zeta _3}{9}-\frac{635 \pi ^4}{36}-\frac{44 \pi^2}{27}
   -\frac{4294}{81}\right)
   +\Bigg[\left(\frac{16}{27}+\frac{1394 \pi ^2}{3}\right) \zeta_3\nn\\
   &\quad-\frac{160552 \zeta _5}{15}+\frac{3247 \pi ^4}{135}
   -\frac{385 \pi ^2}{81}-\frac{35723}{243}\Bigg]\epsilon 
   +\ord{\epsilon^2}
\,,\\
K_{FAf} ={}& -\frac{352}{27 \epsilon ^3}
+\frac{1}{\epsilon ^2}\left(\frac{32 \pi ^2}{27}-\frac{1096}{81}\right)
   +\frac{1}{\epsilon }\Bigg(\frac{640\zeta _3}{9}-\frac{392 \pi ^2}{81}
   -\frac{2000}{81}\Bigg)+\Bigg(\frac{2992 \zeta_3}{27}\nn\\
   &\quad+\frac{44 \pi ^4}{9}-\frac{14398 \pi^2}{243}-\frac{48236}{729}\Bigg)
   +\Bigg[\left(-\frac{2360}{27}-\frac{2128 \pi ^2}{9}\right) \zeta_3+2592 
\zeta _5\nn\\
   &\quad-\frac{2651 \pi ^4}{405}+\frac{154540 \pi ^2}{729}
   -\frac{537494}{2187}\Bigg]\epsilon
   +\ord{\epsilon ^2}
\,,\\
K_{Fff} ={}& \frac{64}{27 \epsilon ^3}+\frac{64}{81 \epsilon ^2}
+\frac{16 \left(\pi ^2-4\right)}{27 \epsilon
   }+\left(-\frac{704 \zeta _3}{27}+\frac{16 \pi ^2}{81}
   -\frac{7744}{729}\right)\nn\\
   &+\left(-\frac{704 \zeta_3}{81}+\frac{134 \pi ^4}{405}-\frac{16 \pi ^2}{27}
   -\frac{70144}{2187}\right) \epsilon
   +\ord{\epsilon ^2}
\,.
\end{align}

\phantomsection
\addcontentsline{toc}{section}{References}
\bibliographystyle{jhep}
\bibliography{HtoLsoft}

\end{document}